\documentclass[twocolumn]{aa}  

\usepackage{subfigure}
\usepackage{graphicx}
\usepackage{amsmath} 
\usepackage{changes}
\usepackage{txfonts}
\usepackage{booktabs}
\usepackage{tikz}
\usetikzlibrary{shapes,arrows,chains,decorations.text}
\usetikzlibrary{arrows,decorations.pathmorphing,backgrounds,
	            positioning,fit,matrix}
\tikzstyle{arrow}=[-latex,thick, ->, >=stealth,latex-latexnew, arrowhead=1cm]
\definecolor{burntorange}{cmyk}{0,0.6,1,0}
\def\oran{orange!30}
\tikzstyle{superpeers}=[draw,circle,burntorange,left 
color=\oran,  text=violet,minimum width=28pt]
\usepackage{caption}
\usepackage[switch,columnwise]{lineno}

\begin{document} 

\title{Atomic carbon, nitrogen, and oxygen forbidden 
emission lines in the water-poor comet C/2016 R2 
(Pan-STARRS)}
  \subtitle{}
  \author{S. Raghuram$^1$, D. Hutsem\'{e}kers$^2$, C. 
  Opitom$^3$, E. Jehin$^2$, A. Bhardwaj$^1$, and J. 
  Manfroid$^2$}
  \institute{$^1$Physical Research Laboratory, Ahmedabad, 
  380009, India. \\
  $^2$STAR Institute - University of Liège, Allée du 6 Août 
  19c,  B-4000 Liège, Belgium. \\
  $^3$ESO (European Southern Observatory) - Alonso de 
  Cordova 3107,  Vitacura, Santiago Chile.}
  \date{}

 
  \abstract
   { {The N$_2$ and CO-rich and water-depleted comet C/2016 R2 
   (Pan-STARRS) (hereafter `C/2016 R2') is a unique comet for 
   detailed 
   spectroscopic analysis.}}
   {We aim to explore the associated photochemistry of 
   parent species, which produces different metastable 
   states and forbidden emissions, in this cometary coma of 
   peculiar composition.}
   { {We re-analyzed the} high-resolution spectra of comet 
   C/2016 R2, 
    {which  {were} obtained} in February 
   2018, 
   using the UVES spectrograph of the European Southern    
   Observatory (ESO) Very Large Telescope (VLT).
    Various  forbidden atomic emission 
   lines of  {[CI], [NI], and [OI]}  {were} observed 
   in the optical    spectrum    of 
   this comet when it was at 2.8 au from the 
   Sun.  The 
   observed forbidden emission intensity ratios are studied 
   in the framework of a couple-chemistry emission model.}
   {The model calculations show that CO$_2$ is the major 
   source of both atomic oxygen green and red-doublet 
   emissions in the coma of C/2016 R2 (while for most comets it 
   is generally H$_2$O), whereas, CO and N$_2$
   govern the atomic carbon and nitrogen  emissions, 
   respectively.  Our modelled oxygen green to red-doublet 
    and carbon to nitrogen  emission ratios 
   are higher by a factor  {of 3}, when 
   compared to the observations. These discrepancies can be 
   due to uncertainties associated with photon cross 
   sections or unknown 
   production/loss sources.  Our modelled oxygen green to red-doublet 
   emission ratio is close to the observations, when we 
   consider an O$_2$ abundance with a production rate of 30\% relative to the 
   CO production rate.  We constrained the mean photodissociation yield 
   of CO producing C($^1$S) as about 1\%, a quantity which 
   has not been measured in the laboratory. 
    {The collisional quenching is not 
   	a significant loss process for N($^2$D) though its radiative lifetime
   	is significant ($\sim$10 hrs). 
   Hence, the observed [NI] doublet-emission ratio 
   ([NI]  5198/5200) of 1.22, which is smaller than the terrestrial measurement 
   by a 
   factor  {1.4}, is mainly due to the characteristic radiative decay of 
   N($^2$D).}}
  {}

   \titlerunning{Forbidden emission lines in the water-poor 
   comet C/2016 R2}
   \authorrunning{Raghuram et al.}
%
   \maketitle

\section{Introduction}

Comets are considered as pristine objects and can provide 
clues about the initial formation conditions of the Solar 
System. These objects have undergone little alteration 
since birth because they spent most of their lifetime in 
remote reservoirs far away from the Sun. Various successive 
ground and spacecraft observations have confirmed that 
comets are primarily composed of water ice amalgamated with 
organic molecules and refractory dust.  When a comet 
reaches the vicinity of the Sun, solar radiation sublimates 
the ices present in the nucleus and produces a gaseous 
envelope called a cometary coma. The interaction of the 
solar radiation with the coma drives a plethora of 
chemical reactions and leads to various spectral emissions 
from cometary species \citep{Feldman04, Rodgers04}.

Remote spectroscopic observations are the primary tool to 
reveal the parent species distribution in the cometary 
comae and also to assess the global composition of comets 
\citep{Feldman04}. Among the various emissions observed in 
cometary spectra,  {the emissions due to optically forbidden 
electronic} transitions of 
metastable species  {have been used as direct 
tracers of cometary parent molecules}. The 
formation of  {these metastable species occurs} mainly due to the 
dissociative excitation of parent species in the coma.  {Solar resonance 
fluorescence is an insignificant excitation 
mechanism in populating these excited states \citep{Festou81}.} 
Hence, the radiative emissions of these species have been 
used as a proxy to trace and also to quantify the 
sublimation rates of parent species in comets  { 
\citep{Delsemme76, Delsemme79, Fink84, Magee90, Schultz92, 
Morgenthaler01, Furusho06, McKay12,McKay12b,McKay19, 
Decock13,Decock15}.}

In spite of being the predominant species in the 
coma of most comets, water can not be observed 
directly in the visible region due to the lack of 
electronic transitions. Hence, atomic oxygen forbidden 
emissions (red-doublet at 6300 \& 6364 \AA, and green at 
5577 \AA\ wavelengths), which are produced during the
photodissociation of water, have been regularly used to 
quantify the 
H$_2$O sublimation rate of the nucleus \citep[see][for more 
details on various atomic oxygen emission observations in 
comets]{Bhardwaj12}. Similarly, the atomic carbon 1931 
\AA\ emission line, which is mainly due to the resonance 
fluorescence of the C($^1$D) state, has been used as a 
diagnostic tool to derive the CO abundance in several 
comets \citep{Feldman76, Smith80, Feldman80, Feldman97, 
Tozzi98}. \cite{Oliversen02} observed C($^1$D) radiative 
decay emissions at wavelengths 9850 and  {9824} \AA\ in comet 
C/1995 O1 Hale-Bopp and the observed emission intensity has 
been used to constrain the CO production rate. 
Even though \cite{Singh91} predicted atomic nitrogen 
forbidden emissions at wavelengths 5200 and 5198 \AA\ in 
comets, these emissions were never observed before. 
\cite{Opitom19} recently observed comet 
C/2016 R2 and
reported the first-ever atomic nitrogen forbidden 
emissions. 
 
Several works have discussed the photochemistry of 
metastable states in water-dominated cometary comae  
\citep{Festou81, Bhardwaj02, Bhardwaj99a, Saxena02, 
Bhardwaj12, Raghuram13, Raghuram14,  Decock15, Raghuram16, 
Cessateur16a, Cessateur16b}. The observed atomic oxygen 
green to red-doublet emission intensity ratio (hereafter 
[OI] G/R ratio) has been used as a proxy to confirm water 
as a dominant source of these emission lines in comets. 
However, the \cite{Bhardwaj12} modelling studies of atomic 
oxygen emission lines in comet C/1996 B2 (Hyakutake) have 
shown that the [OI] G/R  ratio is not a constant in the 
cometary coma and varies as a function of nucleocentric 
distance. These studies also suggested that the [OI] G/R 
ratio is not a proper tool to confirm that water is the 
dominant source of atomic oxygen forbidden emission lines.  
\cite{Decock15} observed atomic oxygen emissions in 
different comets and found that the [OI] G/R ratio varies 
as a function of the nucleocentric distance. By studying the 
photochemistry of these emission lines, the observed [OI] 
G/R ratio has been used to derive CO$_2$ abundances on 
different comets. \cite{Raghuram13} have studied the  
photochemistry of these lines in a very active  comet 
viz., C/1995 O1 Hale-Bopp, and have shown that  CO$_2$ is 
also a potential candidate for the formation of O($^1$S) 
in the cometary coma when its presence is substantial  
(more than 5\% with respect to H$_2$O). This work has 
also shown that photodissociation of CO$_2$ leads to the 
formation of O($^1$S) with higher excess velocities 
compared to the photodissociation of H$_2$O, which could 
be a reason for the observation of the larger width of green 
lines compared to the red-doublet emissions. By considering 
the photochemical production and loss pathways of major 
species, the modelling studies of \cite{Bhardwaj12}, 
\cite{Raghuram13}, and \cite{Raghuram14} explained the 
observed [OI] G/R ratio in different comets observed at 
various heliocentric distances. These studies also 
constrained the photodissociation yield of water producing 
O($^1$S) at Lyman-$\alpha$ wavelength as about 1\% of the 
total absorption cross section.  After detection of 
molecular oxygen in comets 67P-Churyumov-Gerasimenko 
 {\citep{Bieler15,Altwegg19}} and 1P/Halley \citep{Rubin15}, the 
photochemical model developed by \cite{Cessateur16a} has 
shown that  not considering the role of O$_2$ in the 
photochemistry of [OI] emissions leads to an 
underestimation of the CO$_2$ abundance in comets. 
 
All the previously mentioned works aimed to study  
the photochemistry of forbidden emission lines in a water-dominated cometary 
coma. The recent spectroscopic 
observations of C/2016 R2, which were made when the comet was around
2.8 au from the Sun, have shown that this comet has a 
peculiar composition even compared to other comets observed 
 at large heliocentric distances \citep{Cochran18, Biver18, 
Borro18, Wierzchos18, Opitom19, McKay19}. Cometary optical 
spectra  are usually dominated by various resonance 
fluorescence lines of radicals such as OH, NH, CN, C$_2$, 
C$_3$, and NH$_2$. However, the observations of C/2016 R2 
have shown that its coma is rich in N$_2$ and CO and 
depleted of H$_2$O, HCN, and C$_2$, which is unusual and makes 
this comet a unique case among the comets observed so 
far \citep{Biver18, Opitom19, McKay19}.

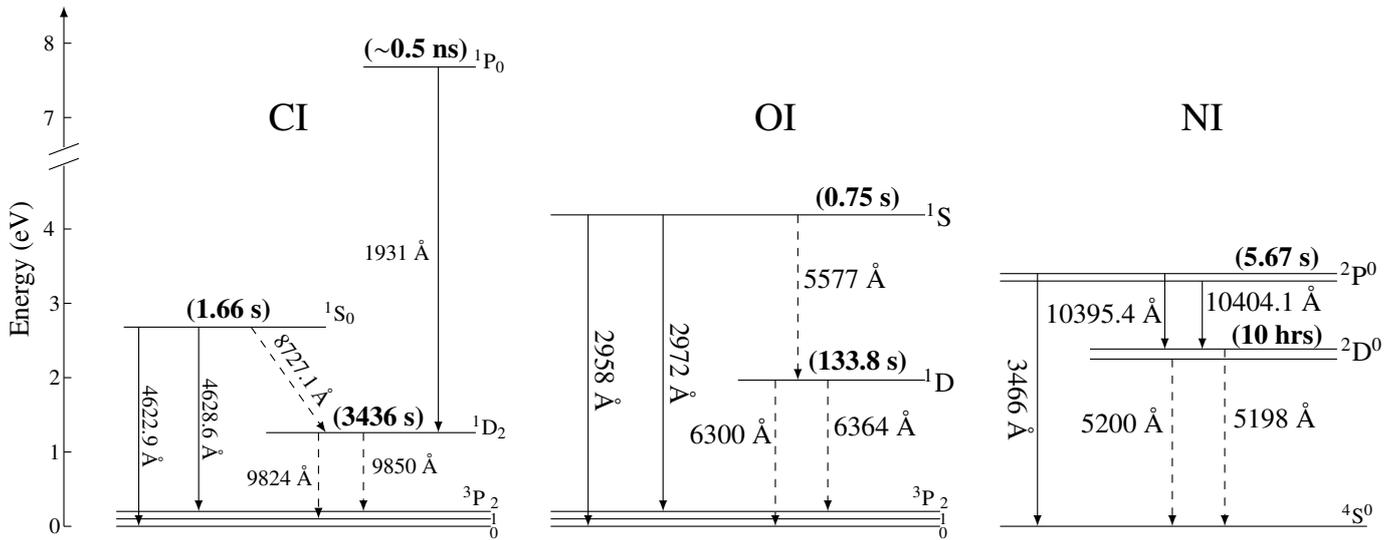
\begin{figure*}[htb]
	\resizebox{\hsize}{!}{
	\begin{tikzpicture} [scale=0.95]
	\begin{scope}[>=latex]
	\draw[-] (0,0) -- (0,4.85) node
	[sloped,midway,yshift=15pt,xshift=28pt] { Energy (eV)};
	\draw[-] (-0.2,4.8) -- (0.2,4.95);
	\draw[-] (-0.2,5) -- (0.2,5.15); 
	\draw[->] (0,5.1) -- (0,7) node
	[sloped,midway,above=3pt] {\hspace{-7cm}};
	
	\draw [-] (0,0) -- (0.1,0) node[at end,xshift = -6pt]{\tiny 0};
	\draw [-] (0,1) -- (0.1,1) node[at end,xshift = -6pt]{\tiny 1};
	\draw [-] (0,2) -- (0.1,2) node[at end,xshift = -6pt]{\tiny 2};
	\draw [-] (0,3) -- (0.1,3) node[at end,xshift = -6pt]{\tiny 3};
	\draw [-] (0,4) -- (0.1,4) node[at end,xshift = -8pt]{\tiny 4};
	\draw [-] (0,5.5) -- (0.1,5.5) node[at end,xshift = -8pt]{\tiny 7};
	\draw [-] (0,6.5) -- (0.1,6.5) node[at end,xshift = -8pt]{\tiny 8};
	
	\node  at (3,5.5) {\Large CI}; 
	
    \draw [-] (0.7,0) -- (5.7,0) node [sloped,at end,above=-8pt]
	{\tiny              \hspace{.1cm}$_0$};
	\draw [-] (0.7,0.1) -- (5.7,0.1) node [sloped,at end]
	{\tiny               \hspace{.1cm}$_1$};
	\draw [-] (0.7,0.2) -- (5.7,0.2) node [sloped,at end,above=-2.5pt]
	{\tiny \hspace{-.3cm}  $^3$P $_2$};

	\draw [-] (2.7,1.26) -- (5.5,1.26) node [sloped,at 
	end,above=-2.5pt,xshift=5pt,yshift=-3pt]
	{\tiny $^1$D$_2$};
	\draw (4.2,1.45) node {\bfseries(3436 s)};
	\draw [-] (0.8,2.68) -- (3.5,2.68) node [sloped,at   
    end,above=-2.5pt,xshift=5pt] {\tiny              
	$^1$S$_0$};
	\draw (2.2,2.9) node {\bfseries(1.66 s)};
	\draw [-] (4,6.18) -- (5.5,6.18) node [sloped,at 
	end,above=-2.5pt,xshift=5pt,yshift=-3pt]
	{\tiny $^1$P$_0$}; 
    \draw (4.7,6.4) node {\bfseries($\sim$0.5 ns)}; 
	
	\draw [->] (1.0,2.68) -- (1.0,0) node 
	[midway,sloped,yshift=5pt]
	{\tiny \color{black}4622.9 \AA};
	\draw [->] (1.8,2.68) -- (1.8,0.2) node 
	[midway,sloped,yshift=5pt]
	{\tiny \color{black}4628.6 \AA};
	\draw [dashed] [->] (3.4,1.26) -- (3.4,0.1) node 
	[midway,xshift=-14pt]
	{\tiny \color{black}9824 \AA};
	\draw [dashed] [->] (4,1.26) -- (4,0.2) node 
	[midway,xshift=15pt,yshift=3pt]
	{\tiny \color{black}9850 \AA};
	\draw [->] (5,6.18) -- (5,1.26) node 
	[midway,xshift=-15pt]
	{\tiny \color{black} 1931 \AA};
	\draw [dashed] [->] (2.5,2.68) -- (3.5,1.26) node 
	[sloped,midway,yshift=7pt]
	{\tiny \color{black} 8727.1  \AA}; 
	
	\node  at (9.5,5.5) {\Large OI}; 
	\draw [-] (6.5,0) -- (11.7,0) node [sloped,at 
	end,above=-8pt]	{\tiny  \hspace{.1cm}$_0$};
	\draw [-] (6.5,0.1) -- (11.7,0.1) node [sloped,at end]
	{\tiny               \hspace{.1cm}$_1$};
	\draw [-] (6.5,0.2) -- (11.7,0.2) node [sloped,at 
	end,above=-2.5pt]
	{\tiny           \hspace{-.3cm}  $^3$P $_2$};
	
	\draw  (9,1.967) -- (11.5,1.967) node [at end, 
	xshift=5pt] {$^1$D};
	\draw (10.6,2.2) node {\bfseries(133.8 s)}; 
	\draw (6.5,4.19) -- (11.5,4.19) node [at end, 
	xshift=5pt] {$^1$S}; %
		\draw (10.6,4.4) node {\bfseries(0.75 s)}; 
	
	\draw [dashed] [->] (9.5,1.967) -- (9.5,0) node 
	[midway,above,xshift=-16pt]
	{\color{black}6300 \AA};   
	\draw [dashed] [->] (10.2,1.967) -- (10.2,0.2) node
	[midway,above,xshift=17pt] {\color{black}6364 \AA};   %
	\draw [dashed] [->]  (9.8,4.19) -- (9.8,1.967) node
	[midway,above,xshift=17pt] {5577 \AA};
	\draw [->] (8,4.19) -- (8,0.2)  node
	[midway,sloped,yshift=7pt] {2972 \AA};  
	\draw [->] (7,4.19) -- (7,0) node
	[midway,sloped,yshift=7pt] {2958 \AA};  
	
	\node  at (15.2,5.5) {\Large NI}; 
	\draw [-] (12.5,0.0) -- (17.4,0.0) node [sloped,at
	end,above=-2.5pt]{\tiny \hspace{-.3cm}  $^4$S$^0$};
	
	\draw  (13.7,2.385) -- (17,2.385) node [at end, 
	xshift=9pt]{$^2$D$^0$};
	\draw  (13.7,2.25) -- (17,2.25) ;  
	\draw (16.2,2.55) node {\bfseries(10 hrs)}; 
	\draw (12.5,3.4) -- (17,3.4) node [at end, xshift=8pt] 
	{$^2$P$^0$}; 
	\draw (12.5,3.3) -- (17,3.3); 
	\draw (16.2,3.6) node {\bfseries(5.67 s)};
	
	\draw [dashed] [->] (14.8,2.25) -- (14.8,0) node 
	[midway,above,xshift=-18pt]
	{\color{black}5200 \AA};   
	\draw [dashed] [->] (15.5,2.385) -- (15.5,0.0) node
	[midway,above,xshift=18pt] {\color{black} 5198
		\AA};   
	\draw [->]  (15.2,3.3) -- (15.2,2.38) node
	[midway,above,xshift=22pt,yshift=-2pt]  
	{10404.1 \AA};
	\draw [->]  (14.7,3.4) -- (14.7,2.38) node
	[midway,above,xshift=-22pt,yshift=-8pt]  
	{10395.4 \AA};
	\draw [->] (13,3.4) -- (13,0)  node
	[midway,xshift=0pt,sloped,below,yshift=2pt] {3466 \AA};
	\end{scope}
	\end{tikzpicture}}
	\caption{Partial Grotrian diagrams of atomic carbon 
    (left), oxygen (centre), and nitrogen (right) showing 
    various allowed and forbidden electronic transitions. 
    The dashed lines represent the observed forbidden 
    emission transitions in C/2016 R2. The values in
     {parentheses} represent the radiative lifetime of 
    the excited state,  {which are taken from \cite{Wiese09} and 
    \cite{Wiese96}}.}
	\label{fig:eleves_diag}
\end{figure*}

\cite{Opitom19} have observed several atomic oxygen (5577, 
6300, \& 6364 \AA), carbon (8727,  {9824}, \& 9850 \AA) and 
nitrogen (5197 \& 5200 \AA) forbidden emission lines 
in C/2016 R2.  {Fig.~\ref{fig:eleves_diag} shows} 
the electronic transitions of these  {forbidden emission} lines.
Since these emissions originate from  metastable states,  which have
long radiative lifetimes  ($\ge$0.5 s), the collisional 
quenching in the inner coma should be accounted for while 
determining the emission intensities. Moreover, the 
formation and destruction of these atomic metastable states 
are associated to several parent species present in 
the coma. The main  production pathways of metastable states 
due to the interaction of solar ultraviolet photons ($h\nu$) and 
suprathermal electrons ($e_{ph}$) with major parent species are : 
\[\begin{array}{l}
h\nu\ \&\ e_{ph} + H_2O  \rightarrow O(^1S, ^1D) + H_2  \\
h\nu\ \&\ e_{ph} + CO_2  \rightarrow C(^1S, ^1D) + O(^1S, 
^1D) + O  \\
h\nu\ \&\ e_{ph} + CO  \rightarrow C(^1S, ^1D) + O(^1S, 
^1D)  \\
h\nu\ \&\ e_{ph} + N_2  \rightarrow N(^2D, ^2P) + N \\
\end{array}\] 

The detection of these various forbidden emissions in a water-poor 
 comet ($<$1\% relative to total gas production rate) has motivated us to study 
 the 
photochemistry of the atomic carbon, oxygen and nitrogen 
metastable states and also their emission processes in 
C/2016 R2. We have
modelled all these emission lines in the framework of earlier 
developed coupled-chemistry emission models. The observations 
of this comet are described in Section~\ref{sec:obs}. The inputs
required to model the observed emission ratios and calculations are
described in Section~\ref{sec:model_ips}. The results of model
calculations are presented in Section~\ref{sec:results} followed by a
discussion in Section~\ref{sec:discuss}. We summarise and conclude 
the present work in Section~\ref{sec:sumary}.

\begin{figure}
	\centering
	\resizebox{\hsize}{!}{\includegraphics{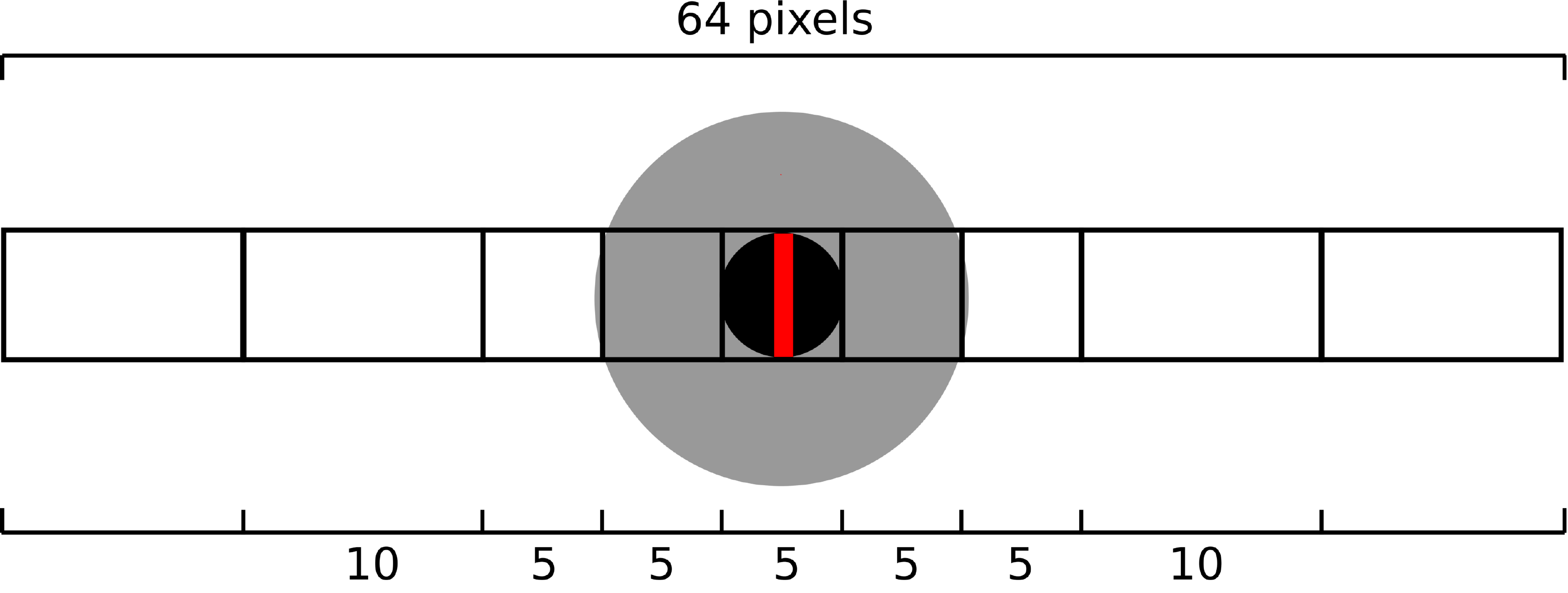}}
	\caption{Slit subdivision. The size of each subslit is
		indicated in pixels. The full slit extends over 64 pixels.
		{The pixel size along the slit is 0.175" on average. 
			Black and gray discs correspond to a seeing disc of 
			0.9" FWHM  and three times the seeing, respectively. 
			The central pixel is denoted by a red rectangle.}
		Adapted from \citet{Decock15}.}
	\label{fig:obs_slit}
\end{figure}

\section{Observations}

\label{sec:obs}


Observations of C/2016 R2 were carried out on 2018 
February 11-16 with the Ultraviolet-Visual Echelle
Spectrograph (UVES) mounted on the 8.2m UT2 telescope of
the European Southern Observatory Very Large Telescope
(VLT). The slit width of 0.44\arcsec\ provides a resolving
power R $\simeq$ 80000. The seeing was 0.9\arcsec\ during
the observations of the [OI], [NI] lines (UVES setting 
390+580) and 1.0\arcsec\ during the observations of the
[CI] lines (UVES setting 437+860).  {More details about}
the observations and data reductions  {are given in}
\citet{Opitom19}.

Surface brightness measurements of the [OI], [CI] and [NI] 
lines were carried out on the two-dimensional spectra, 
cutting the slit in chunks as done in \citet{Decock15} and 
illustrated in Fig.~\ref{fig:obs_slit}. 
 The size of one 
pixel projected onto the comet is 310 km, neglecting the 
differences (smaller than 5\%) between the dates of 
observations and the different settings. The full 
slit thus extends over $\sim$2 $\times$ 10$^4$~km 
projected on the comet. The nights Feb.~13 and Feb.~14 
were affected by thin clouds while the other nights 
were clear.  To account for cloud extinction, the fluxes 
are multiplied by 1.10 and 1.05 for the nights 
Feb.~13 and Feb.~14, respectively. These factors were 
estimated by assuming that the full-slit fluxes in the 
bright lines are identical on Feb.~11, Feb.~13, and 
Feb.~14. The contamination by C$_2$ band emission, which 
generally can affect the [OI]~5577 observation 
\citep{Decock15}, is negligible in this comet due to the very 
small amount of C$_2$ in the coma of this comet \citep{Opitom19}. 
The measurements obtained on the different nights were 
averaged. The extractions done on 
each side of the slit in a given range of nucleocentric 
distances were also averaged. 

The measured surface brightness of the different lines is 
presented in Table~\ref{tab:linebright}, in relative flux 
units, as a function of the nucleocentric projected  
distance.  Errors were computed from propagated photon 
noise and from the dispersion of the night-to-night 
measurements, the largest of the two values being reported 
in  Table~\ref{tab:linebright}. The radius is given as the 
central value in each sub-slit plus or minus the range 
divided by two.

The observed line intensity ratios of interest are given 
in Table~\ref{tab:lineratio}. We  find that at high  nucleocentric projected 
distances ($>$10$^3$
km), the observed [OI]  G/R ratio  
in C/2016 R2 does not match the value of 0.05, which is 
usually measured in water-dominated comets \citep{Decock15}. 
In this Table we also present the observed emission ratio of
[CI] 8727 \AA\ to ( {9824}+9850) (hereafter [CI] ratio)  which is
analogous  to the [OI] G/R 
ratio (see Fig.~\ref{fig:eleves_diag}). However, the [CI] line at 9824 \AA\ was 
contaminated 
by telluric lines so that its surface brightness could not 
be accurately measured  {\citep{Opitom19}}. We thus used the known line 
ratio 
[CI]~9824/[CI]~9850 = 0.34 \citep{Nussbaumer79} to compute 
the [CI] line ratio which is reported in
Table~\ref{tab:lineratio}. Similarly, we have determined
the observed  intensity ratio of [NI] emission lines at  5198 
and 5200 \AA\ ([NI] ratio), which should be equal to  {the} transition 
probabilities ratio of the excited states. Our full
slit averaged [NI] ratio, which is  {1.22 $\pm$ 0.09}, is smaller by a 
factor
of 2.2 than the theoretical branching ratio of 2.7 
\citep[][ {see Section~\ref{sec:doub_rate}}]{Wiese07}, whereas the full 
slit 
averaged [OI]
red-doublet emission ratio of  { 3.03 $\pm$ 0.08} ([OI] ratio, i.e., 
6300/ {6364}) 
is consistent with the theoretically determined 
value \citep{Wiese96}.

The measured and intrinsic widths of the different 
forbidden lines (full width  {at} half maximum, FWHM) are 
presented in Table~\ref{tab:linewidth}. The intrinsic FWHM 
is corrected for the instrumental broadening using the FWHM of 
the ThAr wavelength calibration lines  {using Equation~\ref{eq:fwhm1} and
the obtained intrinsic FWHMs are transformed into 
velocity (km s$^{-1}$) using 
Equation~\ref{eq:fwhm2} that is  
\begin{equation}
FWHM_{intrinsic}(\lambda) = \sqrt{FWHM_{observed}(\lambda)^2 - 
FWHM_{instrumental}(\lambda)^2} 
\label{eq:fwhm1}
\end{equation}

\begin{equation}
FWHM_{intrinsic}(\nu) = \frac{FWHM_{intrinsic}(\lambda)}{\lambda_n\ 2 
\sqrt{ln\ 2}}  
\label{eq:fwhm2}
\end{equation}
where $\lambda_n$ corresponds to the wavelength of forbidden emission 
line. More details about determination of FWHMs are 
given in \cite{Decock13}.}
 {We assume the error on the FWHM$_{instrumental}(\lambda)$ negligible.}
 The FWHMs  {of different spectral emission
lines that are} measured on different 
nights were averaged and the dispersion of the measurements 
used to estimate the error. No significant dependence on 
the nucleocentric distance was found so that only 
measurements for the full slit are reported. The width of 
the green [OI]~5577 line is larger than the width of the 
[OI] red lines, which is similar to the 
observation made in water-dominated comets 
\citep{Decock15}.  {A possible blend affecting [OI] 5577 \AA\ 
was already 
investigated and discarded by \cite{Decock15}}. 
The observed [OI] red lines in 
C/2016 R2 are also wider compared to the values 
obtained in the majority of other comets (2.0 km s$^{-1}$ 
versus $\sim$ 1.5 km s$^{-1}$).    A  {similar} behaviour is
observed for the [CI] lines, with comparable values of the
line widths. On the other hand the faintness of the emission and 
imperfect subtraction of the scattered solar spectrum lead to 
significant errors while determining the [NI] line widths. Hence, 
the difference between the derived widths of the two 
lines is not significant. Given the
uncertainties, the widths of the [NI] lines are  
compatible with the widths of the [OI] and [CI]  lines  {(see Table 
~\ref{tab:linewidth})}.

\begin{table*}[htb]
\caption{Observed surface brightness of the various atomic 
forbidden emission lines in C/2016 R2 as a function of the nucleocentric
projected distance.}
\label{tab:linebright}
\centering
\begin{tabular}{lrrrrrrrrr}
\toprule
 {Distance} ($10^3$ km)   & [OI] 5577  & [OI] 6300  &  [OI] 
6364    & [CI] 
8727     & [CI] 9850     & [NI] 5198     & [NI] 5200    \\
\midrule
0.3875$\pm$0.3875  & 9.82$\pm$0.49 & 26.62$\pm$1.06 & 9.03$\pm$0.45 & 
0.80$\pm$0.12 & 5.13$\pm$0.26 & 1.34$\pm$0.13 & 1.20$\pm$0.12\\
1.5500$\pm$0.7750  & 6.26$\pm$0.31 & 20.81$\pm$0.83 & 6.85$\pm$0.34 & 
0.42$\pm$0.06 & 4.82$\pm$0.24 & 1.24$\pm$0.12 & 0.95$\pm$0.10\\
3.1000$\pm$0.7750  & 3.49$\pm$0.17 & 13.46$\pm$0.54 & 4.49$\pm$0.22 & 
0.35$\pm$0.05 & 4.23$\pm$0.21 & 1.22$\pm$0.12 & 0.92$\pm$0.09\\
5.4250$\pm$1.5500  & 1.92$\pm$0.10 &  8.17$\pm$0.33 & 2.69$\pm$0.13 & 
0.24$\pm$0.04 & 3.01$\pm$0.15 & 0.96$\pm$0.10 & 0.86$\pm$0.09\\
8.4475$\pm$1.4725  & 1.26$\pm$0.06 &  4.96$\pm$0.20 & 1.74$\pm$0.09 & 
0.07$\pm$0.03 & 2.34$\pm$0.12 & 0.91$\pm$0.09 & 0.68$\pm$0.07\\
\midrule
Full Slit          & 3.31$\pm$0.07 & 11.00$\pm$0.22 & 3.63$\pm$0.07 & 
0.35$\pm$0.03 & 3.33$\pm$0.10 & 0.88$\pm$0.03 & 0.72$\pm$0.03\\
\bottomrule
\end{tabular}
\tablefoot{Surface brightness is in arbitrary units}
\end{table*}

\renewcommand{\thefootnote}{\fnsymbol{footnote}}
\begin{table}[htb]
\caption{Observed intensity ratios of atomic 
oxygen, carbon and nitrogen emissions in C/2016 R2 at 
different nucleocentric projected  {distances}.}
\label{tab:lineratio}
\centering
\begin{tabular*}{\columnwidth}{@{\extracolsep{\fill}}lllllcc}
\toprule

 {Distance} ($10^3$ km)            & \small{[OI] $\frac{ 
5577}{(6300+ 6364)}$}  & \small{[CI] $\frac{ 
8727}{(9820+9850)}$\footnotemark[1]} & \small{[NI] $\frac{5198}{5200}$}   \\
\midrule
0.3875$\pm$0.3875  & 0.28$\pm$0.03 & 0.12$\pm$0.02 & 1.12$\pm$0.22\\
1.5500$\pm$0.7750  & 0.23$\pm$0.02 & 0.07$\pm$0.01 & 1.31$\pm$0.26\\
3.1000$\pm$0.7750  & 0.19$\pm$0.02 & 0.06$\pm$0.01 & 1.33$\pm$0.27\\
5.4250$\pm$1.5500  & 0.18$\pm$0.02 & 0.06$\pm$0.01 & 1.12$\pm$0.22\\
8.4475$\pm$1.4725  & 0.19$\pm$0.02 & 0.02$\pm$0.01 & 1.34$\pm$0.27\\
\midrule
Full Slit          & 0.23$\pm$0.01 & 0.08$\pm$0.01 & 1.22$\pm$0.09\\

\bottomrule

\end{tabular*}
\tablefoot{\footnotemark[1]{(9824 + 9850) emission 
intensity is calculated by multiplying the observed 9850 \AA\ 
emission intensity with a factor of 1.34. See the main text for more details.}}
\end{table}

\begin{table}[b]
\caption{Observed line widths of various atomic oxygen, carbon, and 
nitrogen forbidden emissions in C/2016 R2}
\label{tab:linewidth}
\centering
\begin{tabular}{lccc}
\toprule
    & \multicolumn{3}{c}{Full width at half maximum (FWHM)} \\
\cline{2-4} \\[-7pt]
Line  & Observed & Intrinsic    & Intrinsic \\   
(\AA) &  (\AA)   & (\AA)    &  (km s$^{-1}$)   \\     
\midrule		            

{\rm [OI]} 5577   &0.0982$\pm$0.0017  &0.0773$\pm$0.0017  &2.50$\pm$0.05\\
{\rm [OI]} 6300   &0.1019$\pm$0.0010  &0.0702$\pm$0.0010  &2.01$\pm$0.03\\
{\rm [OI]} 6364   &0.1025$\pm$0.0010  &0.0710$\pm$0.0010  &2.01$\pm$0.03\\
{\rm [CI]} 8727   &0.1569$\pm$0.0048  &0.1239$\pm$0.0048  &2.56$\pm$0.10\\
{\rm [CI]} 9850   &0.1611$\pm$0.0023  &0.1069$\pm$0.0023  &1.95$\pm$0.04\\
{\rm [NI]} 5198   &0.0677$\pm$0.0158  &0.0321$\pm$0.0158  &1.11$\pm$0.55\\
{\rm [NI]} 5200   &0.1011$\pm$0.0206  &0.0817$\pm$0.0206  &2.83$\pm$0.72\\
\bottomrule
\end{tabular}
\end{table}

%
%
%

\begin{table*}
	\caption{ {Summary of the baseline input parameters used in the 
	model}}
	\centering
	\begin{tabular}{llll}
		\toprule
		CO gas production rate & Q$_{CO}$ = 1.1 $\times$ 10$^{29}$ s$^{-1}$\\
		Neutral abundances\footnotemark[6] & CO$_2$(18\%), H$_2$O(0.3\%), 
		O$_2$(1\%), and 
		N$_2$(7\%) \\
		Heliocentric distance of the comet & 2.8 au \\
		Geocentric distance of the comet & 2.44 au \\
		Neutral gas expansion velocity & 0.5 km/s\\
		\bottomrule
	\end{tabular}
\tablefoot{\footnotemark[6]{The neutral relative abundances in the parentheses 
are with respect to the CO production rate.}}
\label{tab:baseline}
\end{table*}

\begin{table*}[htb]
	\caption{{Production and loss reactions of  
			N($^2$D), C($^1$D) and C($^1$S) incorporated in the chemical 
			network of coupled-chemistry emission model.}}
	\centering
	\begin{tabular}{lllll}
		\toprule
		Reaction & Rate Coefficient\footnotemark[5] & 
		Reference \\ 
		 & (cm$^3$ molecule$^{-1}$ s$^{-1}$ or s$^{-1}$) & 
		 \\ 			
		\midrule
		h$\nu$ + N$_2$ $\rightarrow$ N($^2$D) + N & 3.0 $\times$ 10$^{-7}$ & 
		This work\\
		h$\nu$ + CN $\rightarrow$ N($^2$D) + C & 1.0 $\times$ 10$^{-6}$ & 
		This work\\
		h$\nu$ + CO $\rightarrow$ C($^1$D) + O & 7.6 $\times$ 10$^{-8}$ & This 
		work\\
		h$\nu$ + CN $\rightarrow$ C($^1$D) + N & 2.0 $\times$ 10$^{-6}$ & 
		This work\\
		h$\nu$ + CO$_2$ $\rightarrow$ C($^1$D) + O & 1.5 $\times$ 10$^{-10}$ & 
        This work\\		
		h$\nu$ + CO $\rightarrow$ C($^1$S) + O & 1.1 $\times$ 10$^{-9}$  & This 
		work\\		
		e$_{ph}$ + N$_2$ $\rightarrow$ N($^2$D) + N & Calculated & 
		This work\\
		e$_{ph}$ + CN $\rightarrow$ N($^2$D) + C & Calculated & 
		This work\\
		e$_{ph}$ + CO $\rightarrow$ C($^1$D) + O & Calculated & This 
		work\\
		e$_{ph}$ + CN $\rightarrow$ C($^1$D) + N & Calculated & 
		This work\\
		e$_{ph}$ + CO$_2$ $\rightarrow$ C($^1$D) + O & Calculated & 
		This 
		work\\		
		N($^2$D) + H$_2$O  $\rightarrow$ N + H$_2$O & 4.0 $\times$ 10$^{-11}$ & 
		\cite{Herron99}\\
		N($^2$D) + CO$_2$  $\rightarrow$ N + H$_2$O & 3.6 $\times$ 10$^{-13}$ & 
		\cite{Herron99}\\
		N($^2$D) + CO  $\rightarrow$ N + H$_2$O & 1.9 $\times$ 10$^{-12}$ & 
		\cite{Herron99}\\
		N($^2$D) + O$_2$  $\rightarrow$ N + O$_2$ & 5.2 $\times$ 10$^{-12}$ & 
		\cite{Herron99}\\
		N($^2$D) + N$_2$  $\rightarrow$ N + N$_2$ & 1.7 $\times$ 10$^{-14}$ & 
		\cite{Herron99}\\ 
		N($^2$D) $\rightarrow$ N + h$\nu_{5200\ \&\ 5198\ \AA}$  & 2.78  
		$\times$ 10$^{-5}$ &  \cite{Wiese09}\\		
		C($^1$D) + H$_2$O  $\rightarrow$ C + H$_2$O & 1.7 $\times$ 10$^{-11}$ & 
		\cite{Schofield79}\\
		C($^1$D) + CO$_2$  $\rightarrow$ C + H$_2$O & 3.7 $\times$ 10$^{-11}$ 
		&\cite{Schofield79}\\
		C($^1$D) + CO  $\rightarrow$ C + H$_2$O & 1.6  $\times$ 10$^{-11}$ & 
		\cite{Schofield79}\\
		C($^1$D) + O$_2$  $\rightarrow$ C + O$_2$ & 2.6  $\times$ 10$^{-11}$ & 
		\cite{Schofield79}\\
		C($^1$D) + N$_2$  $\rightarrow$ C + N$_2$ & 4.2  $\times$ 10$^{-12}$ & 
		\cite{Schofield79}\\
		C($^1$D) $\rightarrow$ C +  h$\nu_{9824\ \&\ 9850\ \AA}$  & 2.91  
        $\times$ 10$^{-4}$ & \cite{Wiese09}\\			
		C($^1$S) + H$_2$O  $\rightarrow$ C + H$_2$O & 1  $\times$ 10$^{-16}$ &
		\cite{Schofield79}\\
		C($^1$S) + CO$_2$  $\rightarrow$ C + H$_2$O & 1  $\times$ 10$^{-16}$ & 
		\cite{Schofield79}\\
		C($^1$S) + CO  $\rightarrow$ C + H$_2$O & 6  $\times$ 10$^{-14}$ & 
		\cite{Schofield79}\\
		C($^1$S) + N$_2$  $\rightarrow$ C + H$_2$O & 3  $\times$ 10$^{-15}$ & 
		\cite{Schofield79}\\
		C($^1$S) $\rightarrow$ C +  h$\nu_{8727\ \AA}$ & 0.6 & 
		 \cite{Wiese09}\\
		\bottomrule
	\end{tabular}
  \tablefoot{ \footnotemark[5]{Photodissociation frequencies presented in this 
  table are calculated at 1 au; e$_{ph}$ is photoelectron and h$\nu$ is solar 
  photon.}}
	\label{tab:rate_coeff}
\end{table*}

\begin{table*}
\caption{Calculated photodissociation frequencies
(s$^{-1}$) of major cometary species producing metastable 
states at 1 au.}
\label{tab:phot_freq}
\centering
	\begin{tabular}{lllllllllll}
	\toprule
Species  &H$_2$O &CO$_2$& CO & O$_2$ & N$_2$ & CN \\
	\midrule	        
	O($^1$S)& 4.2 $\times$ 10$^{-8}$  & 1.0 $\times$ 10$^{-6}$ &  
	4.0 $\times$ 10$^{-8}$\footnotemark[6]  & 1.4 $\times$ 10$^{-7}$ 
	& -- & -- \\ 
	O($^1$D)& 1.0 $\times$ 10$^{-6}$ & 1.6 $\times$ 10$^{-6}$ 
	& 7.6 $\times$ 10$^{-8}$  & 3.6 $\times$ 10$^{-6}$ & -- & --\\
	C($^1$D)\footnotemark[7]& --  & 1.5 $\times$ 10$^{-10}$  
	& 7.6 $\times$ 10$^{-8}$ & -- & -- & 2.0 $\times$ 10$^{-6}$\footnotemark[5] 
	&    \\ 
	N($^2$D)& --  & -- & -- & -- & 3.0 $\times$ 10$^{-7}$  
	& 1.0 $\times$ 10$^{-6}$\footnotemark[1]  \\ 
	\bottomrule
   \end{tabular}
   \tablefoot{ \footnotemark[6]{From \cite{Huebner79}};
   \footnotemark[1]From \cite{Singh91} ;\footnotemark[5]
   This value is calculated using cross section from
   \cite{Qadi13};
   \footnotemark[7]\cite{Huebner92} calculated
   photodissociation frequencies of CH and C$_2$ producing 
   C($^1$D) are 5.1 $\times$ 10$^{-6}$  and 1.0 $\times$ 
   10$^{-7}$, respectively.}
   \label{tab:phrates}
\end{table*}
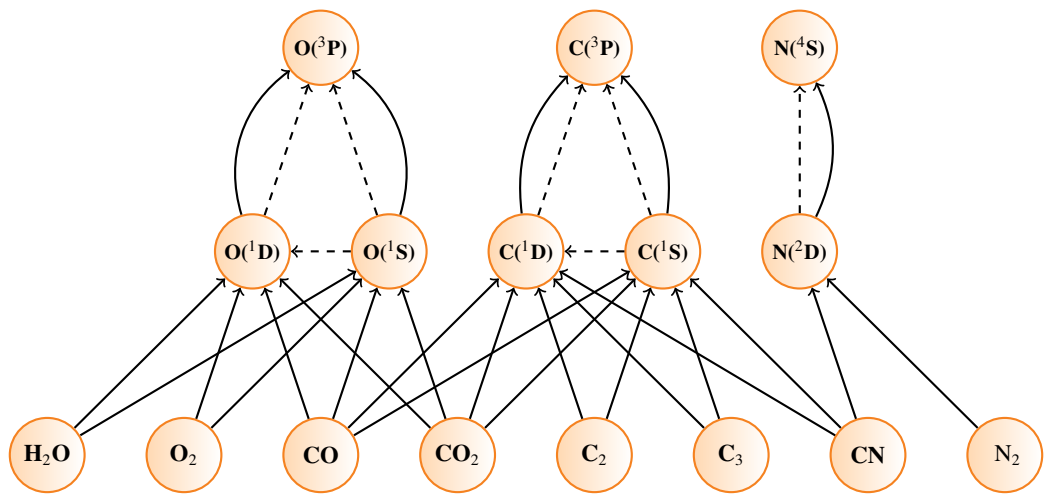
\begin{figure*}
	\centering
	\begin{tikzpicture}[auto, thick, ->,scale=1.8]
	
	\foreach \place/\name in {{(1,0)/a1},
		{(4,0)/a2}, 
		{(3,0)/a3},
		{(2,0)/a4}, 
		{(5,0)/a5}, 
		{(6,0)/a6},
		{(7,0)/a7},                        
		{(8,0)/a8}, 
		%
		{(2.5,1.5)/b1},
		{(3.5,1.5)/b2},
		{(4.5,1.5)/b3},
		{(5.5,1.5)/b4},
		{(6.5,1.5)/b5}, 
		{(3,3)/c1},
		{(5,3)/c2},
		{(6.5,3)/c3}}
	\node[superpeers] (\name) at \place {};
	\node at (1,0) {\small{\bf H$_2$O}}; 
	\node at (4,0) {\small{\bf CO$_2$}}; 
	\node at (3,0) {\small{\bf CO}};     
	\node at (2,0) {\small{\bf O$_2$}};  
	\node at (5,0) {\small{\bf C$_2$}};  
	\node at (6,0) {\small{\bf C$_3$}};  
	\node at (7,0) {\small{\bf CN}};     
	\node at (8,0) {\small{ {N$_2$}}};     
	
	\node at (2.5,1.5) {\tiny{\bf O($^1$D)}}; 
	\node at (3.5,1.5) {\tiny{\bf O($^1$S)}}; 

	\node at (4.5,1.5) {\tiny{\bf C($^1$D)}}; 
	\node at (5.5,1.5) {\tiny{\bf C($^1$S)}}; 
	\node at (6.5,1.5) {\tiny{\bf N($^2$D)}}; 
	
	\node at (3,3) {\tiny{\bf O($^3$P)}}; 
	\node at (5,3) {\tiny{\bf C($^3$P)}}; 
	\node at (6.5,3) {\tiny{\bf N($^4$S)}}; 
	
	\path[every node/.style={font=\sffamily\small}]
	(a1) edge[bend left=0,pos=.55] (b1)
	(a2) edge[bend left=0,pos=.55] (b1)
	(a3) edge[bend left=0,pos=.55] (b1)
	(a4) edge[bend left=0,pos=.55] (b1) 
	
	(a1) edge[bend left=0,pos=.55] (b2)
	(a2) edge[bend left=0,pos=.55] (b2)
	(a3) edge[bend left=0,pos=.55] (b2)
	(a4) edge[bend left=0,pos=.55] (b2) 
	
    (a2) edge[bend left=0,pos=.55] (b3)
    (a3) edge[bend left=0,pos=.55] (b3)
	(a5) edge[bend left=0,pos=.55] (b3)
	(a6) edge[bend left=0,pos=.55] (b3)
	(a7) edge[bend left=0,pos=.55] (b3) 
	
    (a2) edge[bend left=0,pos=.55] (b4)
	(a3) edge[bend left=0,pos=.55] (b4)
	(a5) edge[bend left=0,pos=.55] (b4)
	(a6) edge[bend left=0,pos=.55] (b4)
	(a7) edge[bend left=0,pos=.55] (b4) 
	(a7) edge[bend left=0,pos=.55] (b5) 
	(a8) edge[bend left=0,pos=.55] (b5) 

	(b1) edge[dashed] (c1)     
	(b2) edge[dashed] (c1)
	(b2) edge[dashed] (b1)          
	(b3) edge[dashed] (c2)     
	(b4) edge[dashed] (c2)     
	(b4) edge[dashed] (b3)     
	(b5) edge[dashed] (c3)     
	(b1) edge[bend left=35,pos=0.2] node[midway=-0.5] {} 
	(c1)          
	(b2) edge[bend right=35,pos=0.5] node[above=-0.1] {} 
	(c1)          
	(b3) edge[bend left=25,pos=0.5] node[below=-0.05] {} 
	(c2)          
	(b4) edge[bend right=25,pos=0.5] node[above=-0.05] {} 
	(c2)               
	(b5) edge[bend right=25,pos=0.5] node[above=-0.05] {} 
	(c3)               
	;
	\end{tikzpicture} 
	\caption{ {Schematic presentation of the various 
	chemical pathways of  production and destruction of [CI], 
	[OI], and [NI] metastable states in the cometary coma. 
	Solid arrows, dashed arrows, and solid curved arrows, 
	represent the production channels of the excited species 
	via photon and electron impact dissociative excitation 
	of cometary neutrals, loss of excited species via
	radiative decay, and loss of excited species due to 
	collisional quenching with neutrals, respectively. In the non-collisional 
    region of the coma, all these photochemical reactions, except the
    collisional quenching reactions (curved arrows), determine the intensities 
    of forbidden emission lines.}}
    \label{fig:photpaths}	
\end{figure*}

\section{Model calculations}
\label{sec:model_ips}
The model calculations of the atomic oxygen green and 
red-doublet emissions intensities are explained in detail 
in our earlier work \citep{Bhardwaj12, 
	Raghuram13, Raghuram14, Decock15, Raghuram16}. We have 
updated our model by including the atomic carbon and nitrogen 
emission lines. Here we present the model inputs considered 
for C/2016 R2 for the observational conditions of 
\cite{Opitom19}. We have done the model 
calculations  {when the comet was at} distances of 
2.8 au from the Sun and 2.44 au from the Earth. 
Since there is no information available on the radius of 
this comet, we assumed a typical value of 10 km for an Oort cloud comet.

\subsection{ {Neutral distribution}}
The primary neutral composition of the cometary coma is taken as CO, 
CO$_2$, H$_2$O, N$_2$, and O$_2$. We have taken a CO gas production 
rate for this comet of 1.1 $\times$ 10$^{29}$ s$^{-1}$  
\citep{Biver18}. The relative abundances of CO$_2$ and 
H$_2$O with respect to CO are taken as 18\% and 0.3\%, 
respectively, from  \cite{McKay19}.  {Based on the measured ionic 
emission intensity ratios}, the relative 
volume mixing ratio of N$_2$ is taken as 7\% with 
respect to CO 
\citep{Cochran18, Opitom19, McKay19}. 
The recent Rosetta mass spectrometer in-situ measurements 
on 67P/Churyumov-Gerasimenko  {\citep{Bieler15,Altwegg19}}
and subsequent investigation of data of the Giotto mass 
spectrometer from comet 1P/Halley \citep{Rubin15} have 
suggested that O$_2$ might be a common and an abundant 
primary species in comets. In order to incorporate the
photodissociation of O$_2$ in the model, we have assumed
an abundance of 1\% with respect 
to CO. However, we vary this value to study its impact on 
the calculations of the [OI] G/R ratio.  {We adopted 
an outflow flow velocity of the neutral gas as 0.85 
$\times {r_h}^{-0.5}$  = 0.5 km/s at heliocentric distance $r_h$ 
\citep{Cochran93,deleion19},} and spherical symmetry is considered in the 
calculations.

 {The model calculates the neutral density 
profiles of the primary cometary species using the following Haser's formula, 
which 
assumes spherical expansion of volatiles into the space with a constant 
radial expansion velocity 
\citep{Haser57}.} 
\begin{equation}
    n_i(r) = \frac{f_i Q_0}{4 \pi r^2 v_i} exp(-\beta_i r)
\end{equation}
 {Here  $n_i(r)$ is the number density of the $ith$ neutral species at 
radial distance $r$  and $Q_0$ is the total gas production rate of the 
comet. $v_i$, $f_i$, and $\beta_i$ represent the expansion velocity, 
fractional composition, and inverse of scale height of the $ith$ neutral 
species. We have considered N$_2$ {and CN} as the primary species to
produce N($^2$D) via dissociative excitation by both photons and 
photoelectrons. Assuming HCN is the 
parent source of CN, the neutral density profile of CN is calculated using 
Haser's two-component formula \citep{Haser57}. For this calculation, we have 
taken HCN gas production as 4 $\times$ 10$^{24}$ s$^{-1}$ from the 
measurements of \cite{Biver18}. The role of {other} 
minor N-bearing  species in producing N($^2$D) is found to be
negligible as discussed later.}

Aforementioned cometary gas parameters, which resemble the 
observational conditions of C/2016 R2, are taken as baseline 
inputs for the model calculations.  {These baseline 
	parameters are summarized in Table~\ref{tab:baseline}.} However, 
we vary the neutral abundances used in the model to discuss the effect 
on the calculated emission ratios. 

\subsection{ {Atomic and molecular parameters}}
\label{sec:atom_prop}
 {A detailed explanation of the input cross sections and chemical 
network, which 
have been used to calculate [OI] forbidden emission lines, can be found in our 
earlier modelling work \citep{Bhardwaj12}.  These model calculations   
account for various production and loss mechanisms of atomic oxygen in $^1$S 
and $^1$D states. We have added various 
production and loss reactions of metastable C($^1$D), C($^1$S) and N($^2$D) 
in the [OI] chemical network to calculate the atomic carbon and nitrogen  
emission line intensities. Table~\ref{tab:rate_coeff} shows 
these additional set of reactions, which are  incorporated in our new 	
chemical network.} 
	
We have taken the electron impact cross	section of N$_2$  producing N($^2$D) 
from \cite{Tabata06}. To incorporate production of N($^2$D) via 
photodissociation of N$_2$, following the approach of \cite{Fox93}, we have
assumed an average photodissociation yield of 0.32. The impact of this 
assumption on the calculation of [CI]/[NI] emission ratio will be discussed in 
the later section.   {We have taken the photodissociation cross
sections of CN producing C($^1$D) and N($^2$D) from \cite{Qadi13}.}
The photodissociation cross section of CO producing C($^1$S) has never been 
reported in the literature. In 	order to incorporate this production path in 
the model, we assumed that  0.5\% of the total CO absorption cross section, 
above the dissociation threshold energy, leads to C($^1$S) formation. However, 
we vary this value 	to fit the observed carbon emission intensity ratios, 
as discussed later.  {The 
collisional quenching rate coefficients of N($^2$D) by H$_2$O, CO, CO$_2$, and 
O$_2$ are taken from \cite{Herron99}, whereas, for C($^1$D) they are 
taken from \cite{Schofield79}.}

\subsection{ {Determination of emission intensity ratios}}
 {The solar radiation flux  {($\phi_{ph}(\lambda)$)} in the 
wavelength region 5--1900 \AA\ is degraded in the cometary coma using the
Beer-Lambert's law. The 
photoelectron production rate spectrum is calculated as a function of 
electron energy at different radial distances. By incorporating the Analytical 
Yield Spectrum 
(AYS) 
approach, which accounts for the degradation of electrons in the cometary coma 
based on the Monte Carlo technique, the suprathermal flux ($\phi_e(r,E)$) is 
calculated as a function of electron energy (E)  {at the radial distance 
r}. 
More 
details about the calculation
of suprathermal electron flux are given in \cite{Bhardwaj12}.}

 {By using the model degraded photon 
flux and steady state suprathermal electron flux profiles with the 
corresponding  photon and electron impact excitation cross sections, we 
determined the  {volume} production rate profiles for the different 
metastable 
state 
species in the coma  {using the following equations.}
\begin{equation}
P_{ij,photon}(r) = n_i(r) \int_{\lambda_{min}}^{\lambda_{max}}  
\sigma_{ij, photon}(\lambda)\ 
\phi_{ph}(\lambda)\ e^{-\tau(\lambda,r)} \ d\lambda
\end{equation}

\begin{equation}
P_{ij,electron}(r) = n_i(r) \int_{E_{min}}^{E_{max}}  \sigma_{ij,electron}(E)\ 
\phi_e(r,E) \ dE
\end{equation}
 {Where $P_{ij,photon}(r)$ and $P_{ij,electron}(r)$ are the volume 
production rates of the $jth$ metastable species due to the photon and electron 
impact dissociative excitation of the $ith$ cometary neutral species, 
respectively. 
$\sigma_{ij, photon}$ and  $\sigma_{ij, electron}$
are the respective photon and electron impact dissociative excitation cross 
sections of the $ith$ neutral species producing the $jth$ metastable state. 
$\lambda_{min}$ = 5 \AA\ ($E_{min}$ = 0 eV) and $\lambda_{max}$ = 1900 \AA\ 
($E_{max}$ = 100 eV) are the 
respective lower and upper 
integration limits of the wavelength (energy). $\tau(\lambda,r)$ is the optical 
depth of the medium for the photons of wavelength $\lambda$ at the radial 
distance 
r.}
 Table~\ref{tab:phrates} shows the {calculated} 
photodissociation  frequencies of the major cometary volatiles producing 
different metastable species at 1 au.}
 {The loss rate profiles of these excited states, which are 
mainly due to collisional quenching and radiative decay, are determined by
incorporating the collisional chemistry in the coma. The major photochemical 
pathways, producing the forbidden emission lines, are schematically presented 
in Fig.~\ref{fig:photpaths}.}

 {The calculated production and loss rate profiles have been used to 
obtain the   
number densities of metastable species ($n_j$) by solving the following 
continuity 
equation.}
\begin{equation}
\frac{1}{r^2}\frac{\partial(r^2 n_j v_j)}{\partial r} = 
 P_{j} -  L_{j}  
 \label{eq:conti}
\end{equation}

 {Here $r$ is the radial distance from the cometary nucleus, $v_j$
is the radial transport velocity, and $P_{j}$ and $L_{j}$ are the 
respective total production and loss rates of the $jth$ metastable 
species.} The transport velocities of the metastable species 
are taken from the measured intrinsic line widths 
(see Table~\ref{tab:linewidth}).
The calculated density profiles are converted into volume 
emission rates by multiplying the corresponding Einstein 
 transition probabilities of the different emissions. 
The volume emission rate profiles are integrated along the 
line of sight to obtain surface brightness profiles. We 
have also accounted for the optical seeing effect in 
determining the emission ratio as described in 
\cite{Decock15}. As explained in the earlier section,  
seeing during the observations was 0.9" for [OI], [NI] 
emissions and 1.0" for [CI] emissions.

We have determined the [OI] G/R ratio and emission 
intensity ratios of C($^1$D) to N($^2$D) (hereafter 
[CI]/[NI] ratio), and carbon lines (hereafter [CI] ratio) 
using the following equations. 

\begin{align}
[OI]\ G/R\ ratio  = \frac{ I_{5577\ \AA}}{ I_{6300 \AA} + 
I_{6364 \AA}} = \frac{A_1\ [O(^1S)]}{A_2\ [O(^1D)]} \\
[CI]/[NI]\ ratio = \frac{I_{9850 \AA} + I_{9824 
		\AA}}{I_{5200 \AA} + I_{5198 
		\AA}} = \frac{A_3\ [C(^1D)]}{A_4\ [N(^2D)]}  \\
[CI]\ ratio = \frac{I_{8727 \AA}}{I_{9850 \AA} + I_{9824 
		\AA}} = \frac{A_5\ [C(^1S)]}{A_3\ [C(^1D)]}  
\end{align}

Where A$_1$ (1.26 s$^{-1}$),  { A$_2$ (7.48 $\times$ 10$^{-3}$ 
s$^{-1}$), A$_3$ (2.91 $\times$ 10$^{-4}$ s$^{-1}$),  A$_4$ 
(2.79 $\times$ 10$^{-5}$ s$^{-1}$)}, and A$_5$ (0.6 
s$^{-1}$) are the total Einstein transition probabilities
for the radiative decay of metastable states,
 {which are taken from \cite{Wiese09} and \cite{Wiese96}}. 
[O($^1$S)], [O($^1$D)], [C($^1$D)], [N($^2$D)]  and  [C($^1$S)]
are the number densities.

We have also  {studied} the sensitivity of the model
calculated emission 
ratios  {to several parameters:} the uncertainties 
associated with cross sections,  {neutral abundances, transport velocities 
of the species,} and the 
collisional rate 
coefficients, as discussed in later sections. 

 \section{Results}
 \label{sec:results}
\begin{figure}[htb]
	\centering
	\resizebox{\hsize}{!}{
    \includegraphics{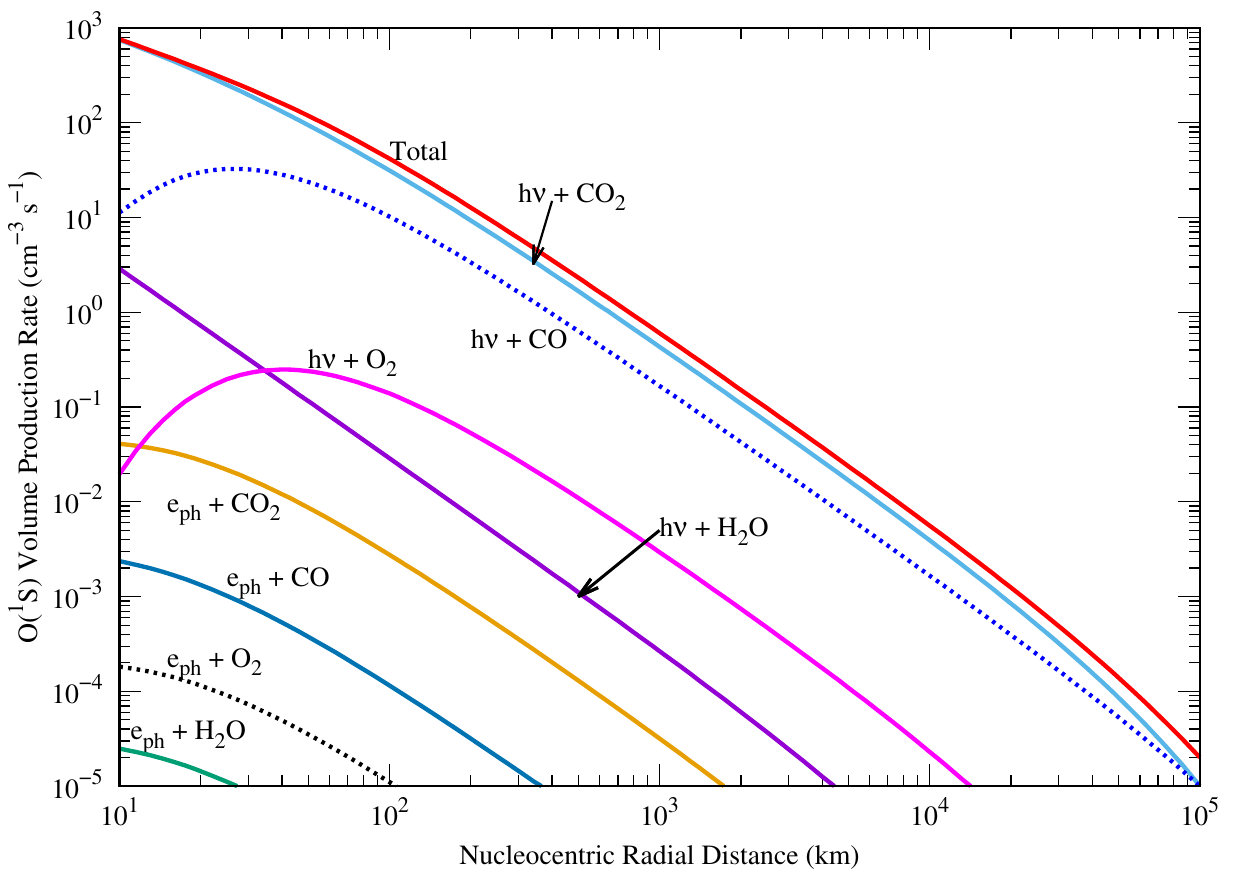}}
    \resizebox{\hsize}{!}{
	\includegraphics{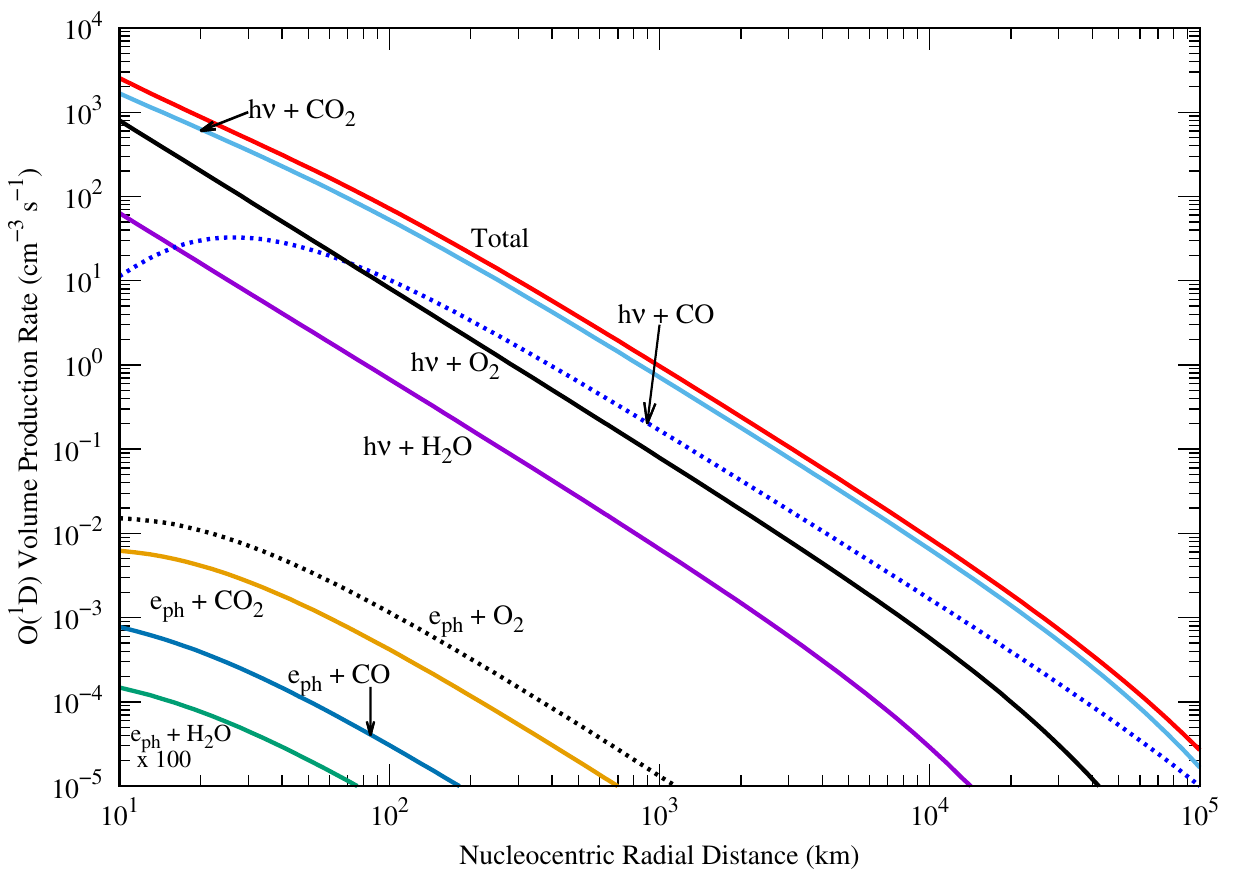}} 
	\caption{Calculated O($^1$S) (top panel) and 
	O($^1$D) (bottom panel) production rate profiles, for 
    different photon and electron impact dissociative 
    excitation reactions of major cometary volatiles, in C/2016 R2 
    when it was at 2.8 au from the Sun. The total gas 
	production rate of CO is taken as 1.1 $\times$ 10$^{29}$
	s$^{-1}$ with 0.3\% H$_2$O,  {18}\% CO$_2$ and 1\% O$_2$ with
	respect to carbon monoxide. h$\nu$ represents a 
	solar photon and e$_{ph}$ is a 	suprathermal electron.}
	\label{fig:pr_rate_o1sd}%
\end{figure}

 The calculated production rate profiles of 
 atomic oxygen in $^1$S (top panel) and $^1$D (bottom 
 panel) states are presented in Fig.~\ref{fig:pr_rate_o1sd}. This 
 calculation  shows that the formation of both O($^1$S) and 
 O($^1$D) in the coma  of C/2016 R2 occurs  mainly due to 
 the photodissociation of CO$_2$. The photodissociation of CO 
 contributes to about 20\% of the total O($^1$S) production
 rate. The contribution of the photodissociative excitation of 
 other oxygen-bearing species  to the total O($^1$D) production
 rate is small  ($<$10\%), within the observed projected
 distance, i.e.  10$^4$ km. The electron impact dissociative 
 excitation of oxygen-bearing species plays a minor role in 
 producing these atomic oxygen  metastable states.

\begin{figure}
	\centering
	\resizebox{\hsize}{!}{
	\includegraphics{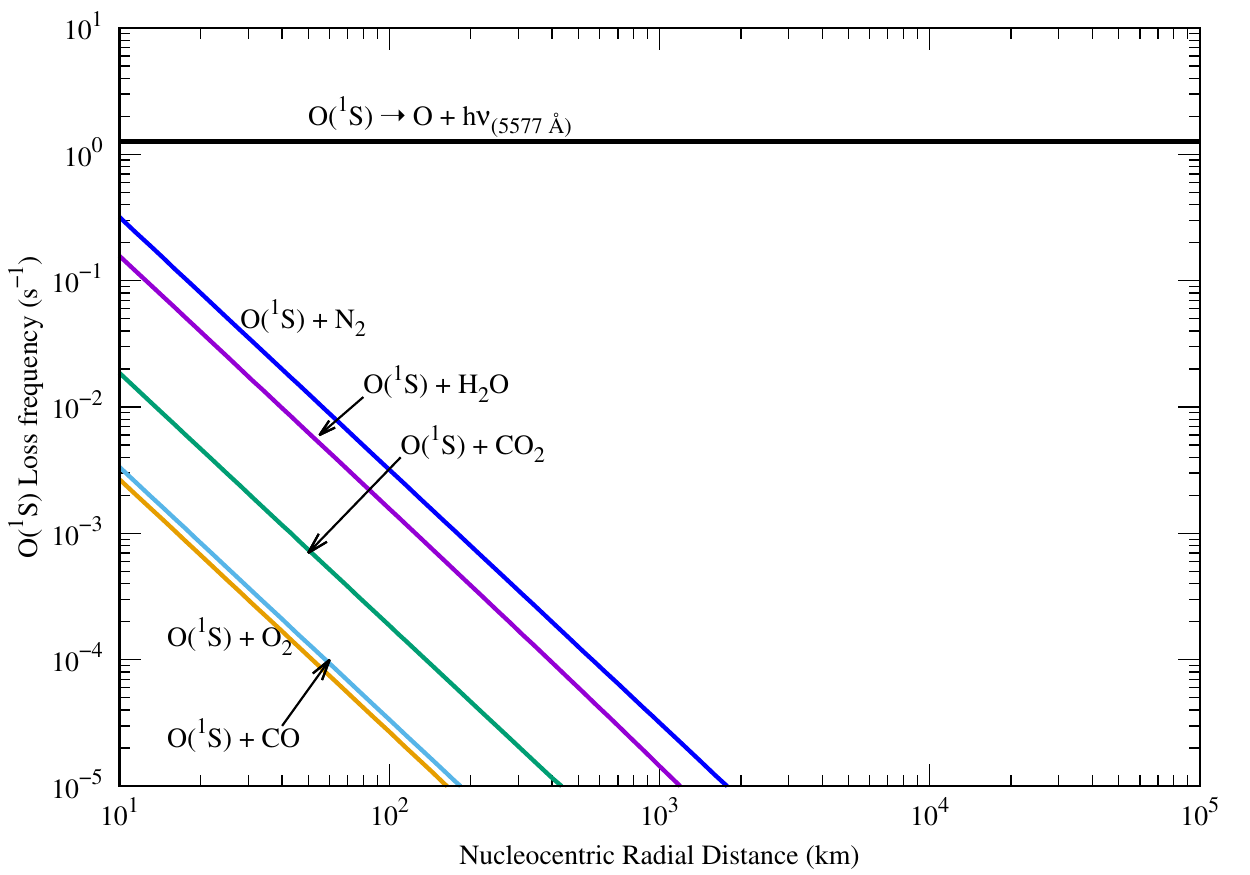}}
	\resizebox{\hsize}{!}{
	\includegraphics{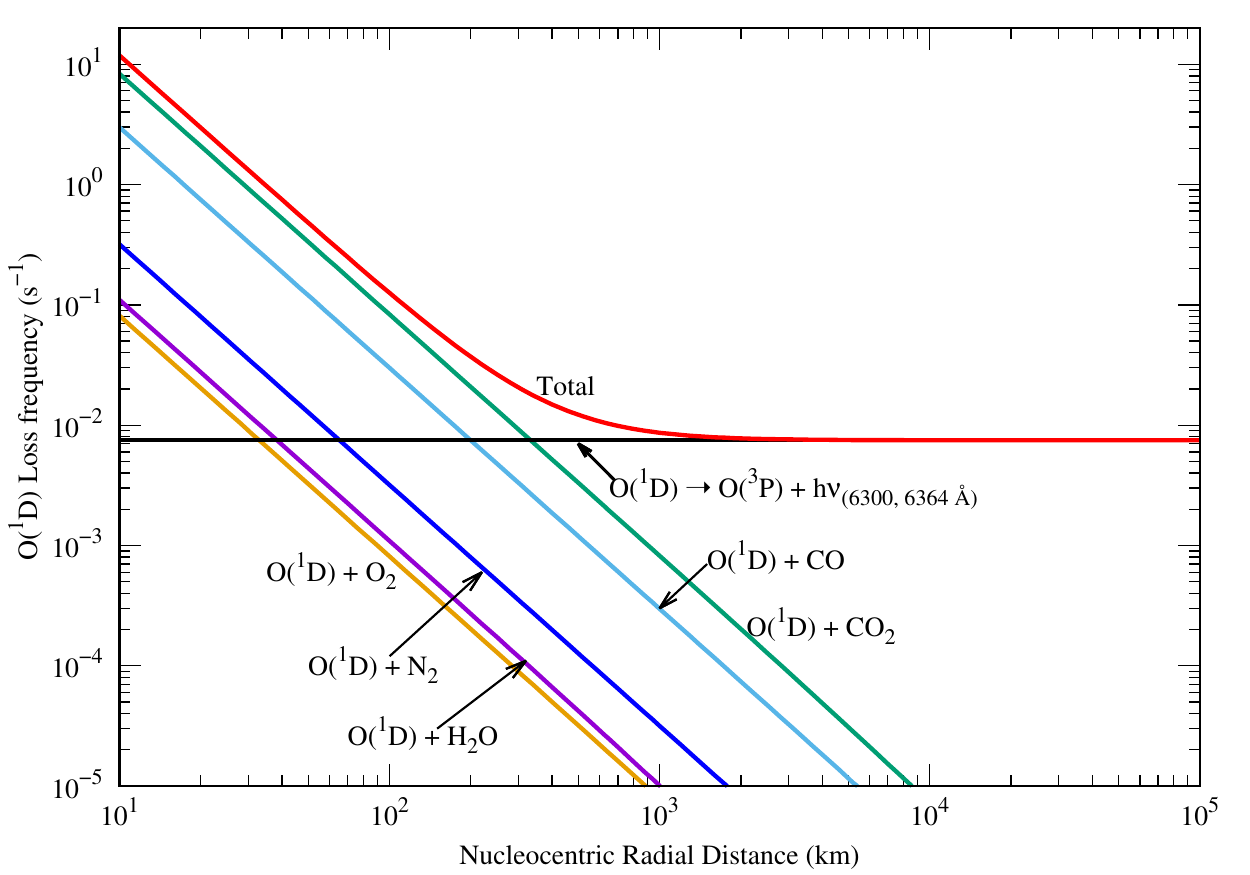}}
	\caption{Modelled O($^1$S) (top panel) and 
	O($^1$D) (bottom panel) loss frequency profiles via 
	collisional quenching of the major cometary volatiles
	and radiative decay 
	in C/2016 R2. 
	The input conditions are the same as in 
	Fig.~\ref{fig:pr_rate_o1sd}.}
	\label{fig:ls_rate_o1sd}%
\end{figure}

The calculated loss frequency profiles of atomic oxygen in 
$^1$S (top panel) and  $^1$D (bottom panel) states are 
presented in Fig.~\ref{fig:ls_rate_o1sd}. Radiative 
decay is the primary loss source of O($^1$S)
in the entire coma, as compared to the collisional 
quenching. Whereas, the CO$_2$ collisional  quenching  is 
the significant loss process for O($^1$D) for radial 
distances below 500 km. Above this radial distance, radiative 
decay is the prominent loss source.

\begin{figure}
	\centering
	\resizebox{\hsize}{!}{
	\includegraphics[scale=0.73]{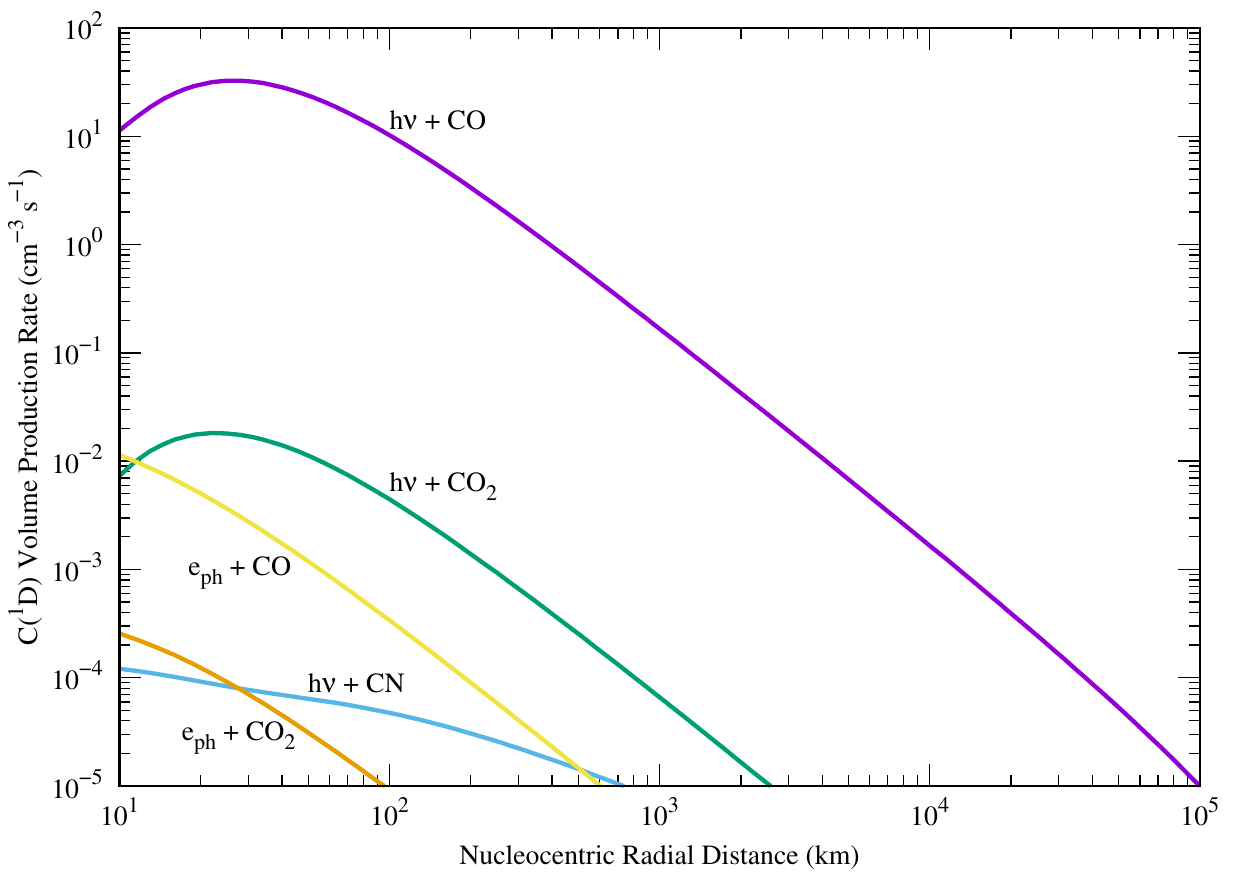}}
	\resizebox{\hsize}{!}{
	\includegraphics[scale=0.75]{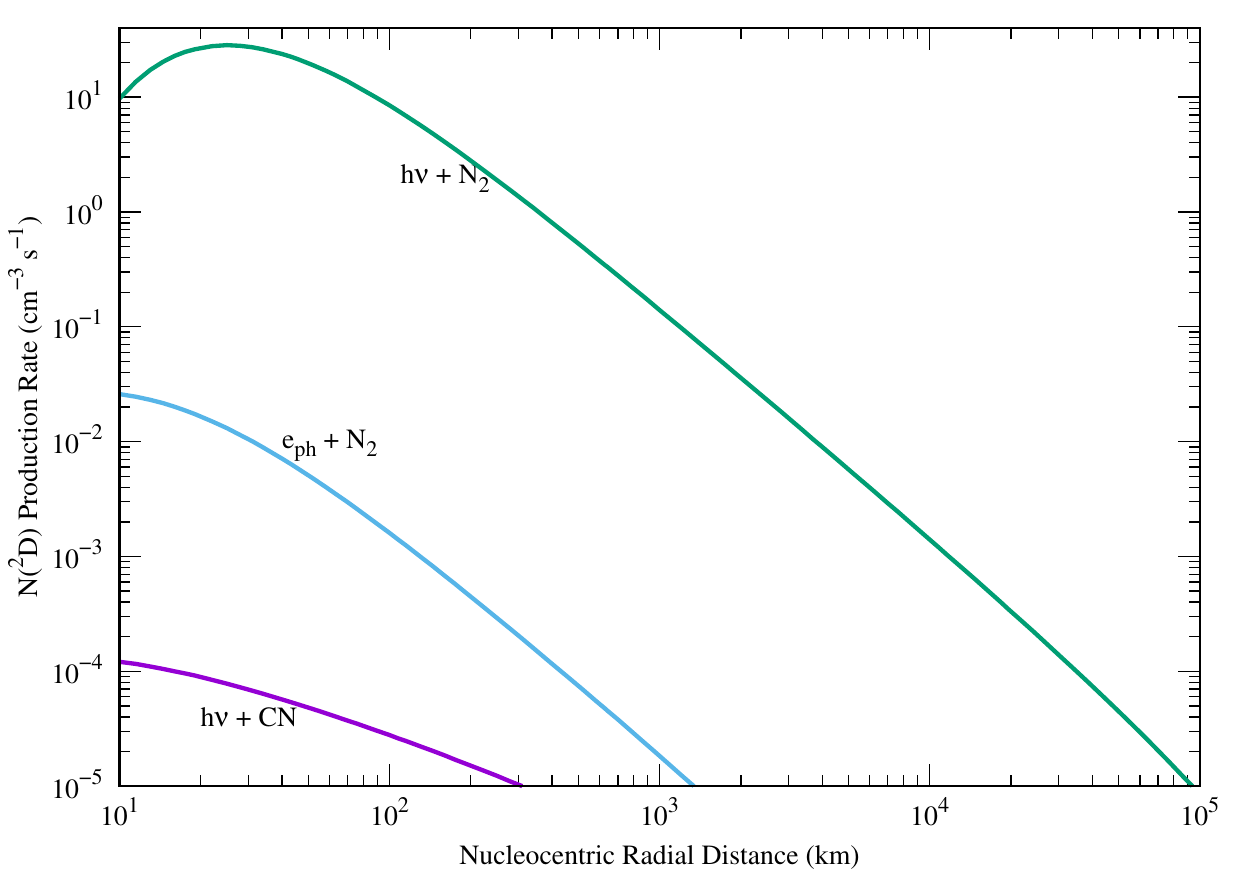}}
	\caption{Calculated C($^1$D) (top panel) and 
	N($^2$D) (bottom panel) production rate profiles, for  
	different photon and electron impact dissociative 
	excitation reactions of the major cometary volatiles, in C/2016 R2.  
	 {Photodissociation and electron impact dissociation of CN
	producing N($^2$D) production rate profiles are plotted after
    multiplying them by a factor 10.}
	The input conditions are the same as explained in 
	Fig.~\ref{fig:pr_rate_o1sd}. h$\nu$ represents a solar 
	photon and e$_{ph}$ is a suprathermal electron.}
	\label{fig:pr_rate_c1dn2d}%
\end{figure}

\begin{figure}
	\centering
	\resizebox{\hsize}{!}{
 	\includegraphics{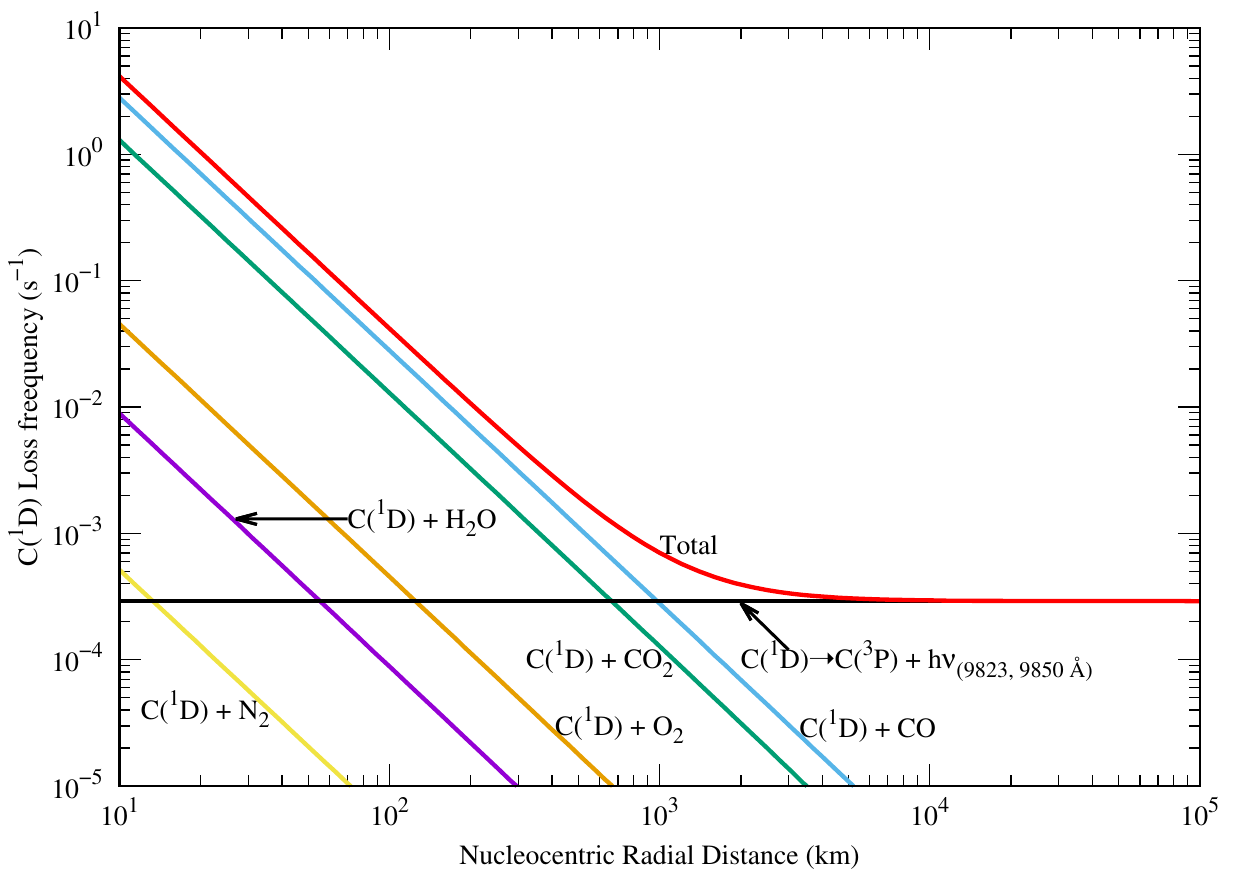}}
	\resizebox{\hsize}{!}{ 
	\includegraphics{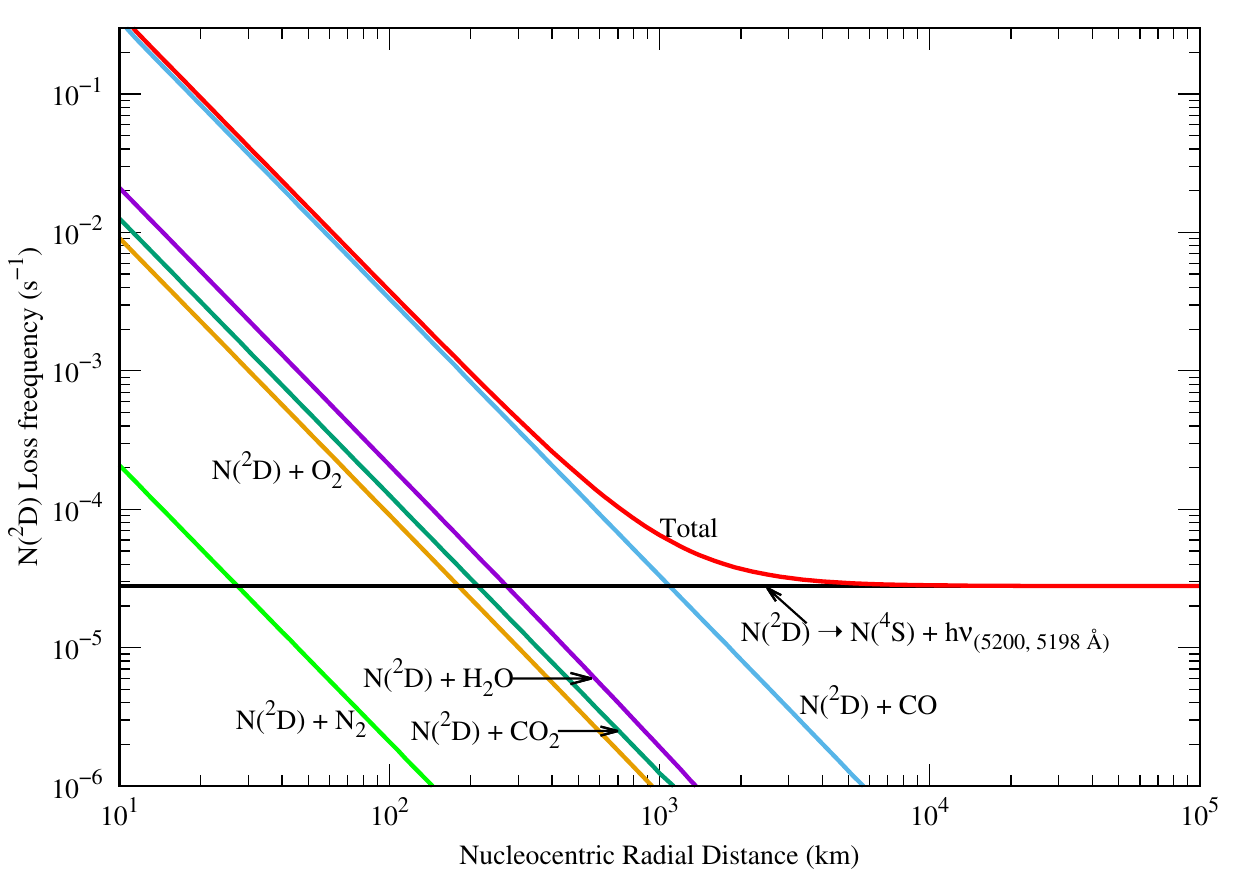}}	
	\caption{Calculated C($^1$D) (top panel) and 
	N($^2$D) (bottom panel) loss frequency profiles via 
	collisional quenching of major cometary volatiles
	 {and radiative decay in C/2016 R2.} 
	The input conditions are the same as in 
	Fig.~\ref{fig:pr_rate_o1sd}.}	
	\label{fig:ls_rate_c1n2d}%
\end{figure}

The modelled  C($^1$D) and N($^2$D) production 
rate profiles for different formation pathways are presented 
in Fig.~\ref{fig:pr_rate_c1dn2d}. This calculation 
shows that the photodissociation of CO and N$_2$ are the 
major formation processes of C($^1$D) and N($^2$D), 
respectively,  whereas the contribution of the 
electron impact dissociative excitation to the total production
rate is negligible. The role of the CO$_2$ photodissociation  is
minor (more than three orders of  magnitude smaller than the CO 
photodissociation) in the total formation of C($^1$D).  {Similarly, 
these calculations also show that the contribution from the
dissociative excitation of CN, in producing C($^1$D) and N($^2$D), is 
negligible 
to the total formation rate.} 
Modelled loss frequency profiles of these excited 
states are presented in Fig.~\ref{fig:ls_rate_c1n2d} and 
show that significant  collisional 
quenching of C($^1$D) and N($^2$D) occurs via CO ($>$90\% of the total), for 
the radial distances below 3 $\times$ 10$^{3}$ km. Above 
this radial distance, radiative decay 
is the major loss mechanism for both C($^1$D) and N($^2$D).

\begin{figure}
	\centering
	\resizebox{\hsize}{!}{	
	\includegraphics{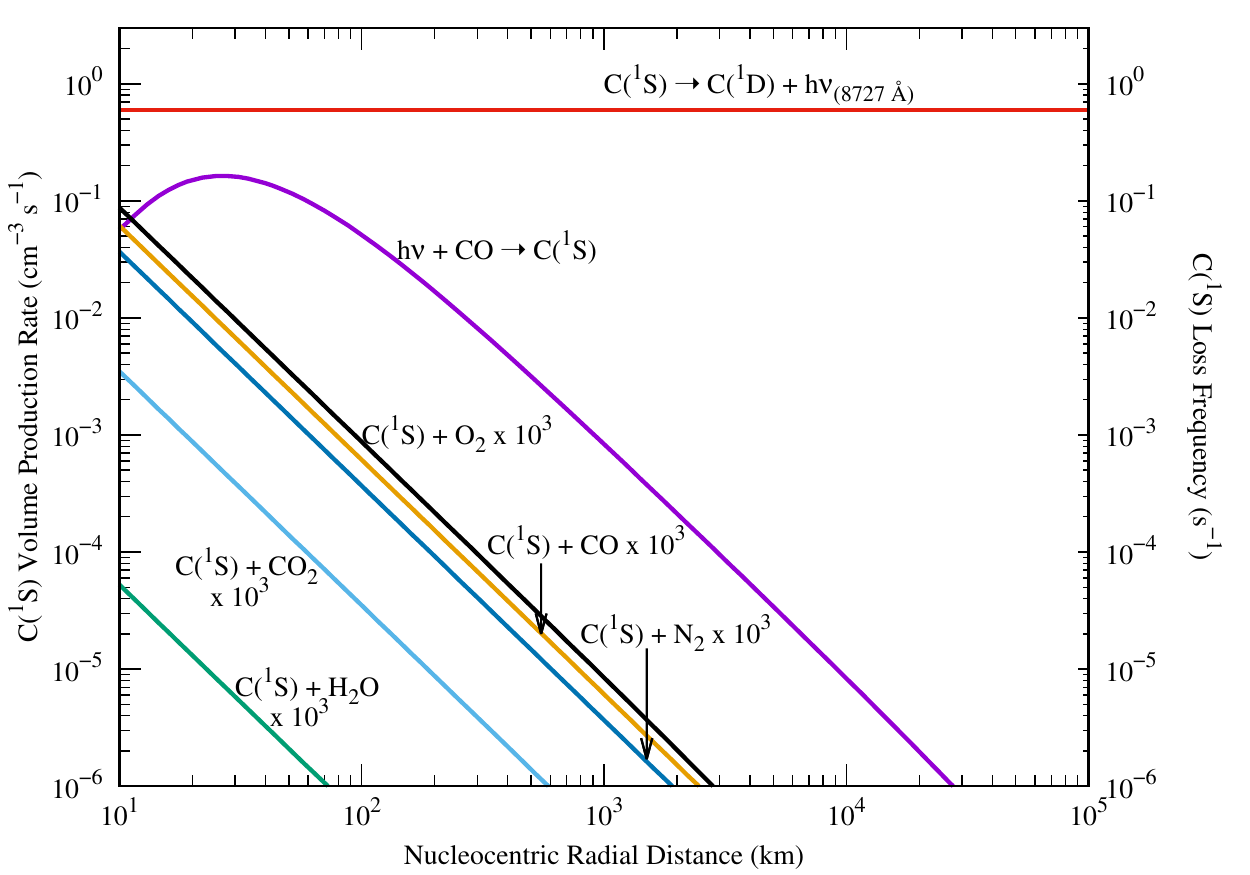}}
	\caption{ {Calculated production rate (left y-axis) and 
		loss frequency (right y-axis)  profiles of C($^1$S) in C/2016 R2.
		The collisional quenching rates are 
		plotted after multiplying them by a factor of 10$^3$.
		The input conditions are the same as in 
		Fig.~\ref{fig:pr_rate_o1sd}.}}	
	\label{fig:ls_rate_c1s}%
\end{figure}

 {The modelled C($^1$S) production rate profile, via 
photodissociation of CO, and the loss profiles due to collisional 
quenching and radiative decay are presented in 
Fig.~\ref{fig:ls_rate_c1s}. Radiative decay of  C($^1$S) 
is found to be the dominant loss channel, which leads to 
[CI] 8727 \AA\ emission, with an  efficiency
several orders of magnitude higher compared to the 
collisional quenching.}

\begin{figure}
	\centering
	\resizebox{\hsize}{!}{	
 	\includegraphics{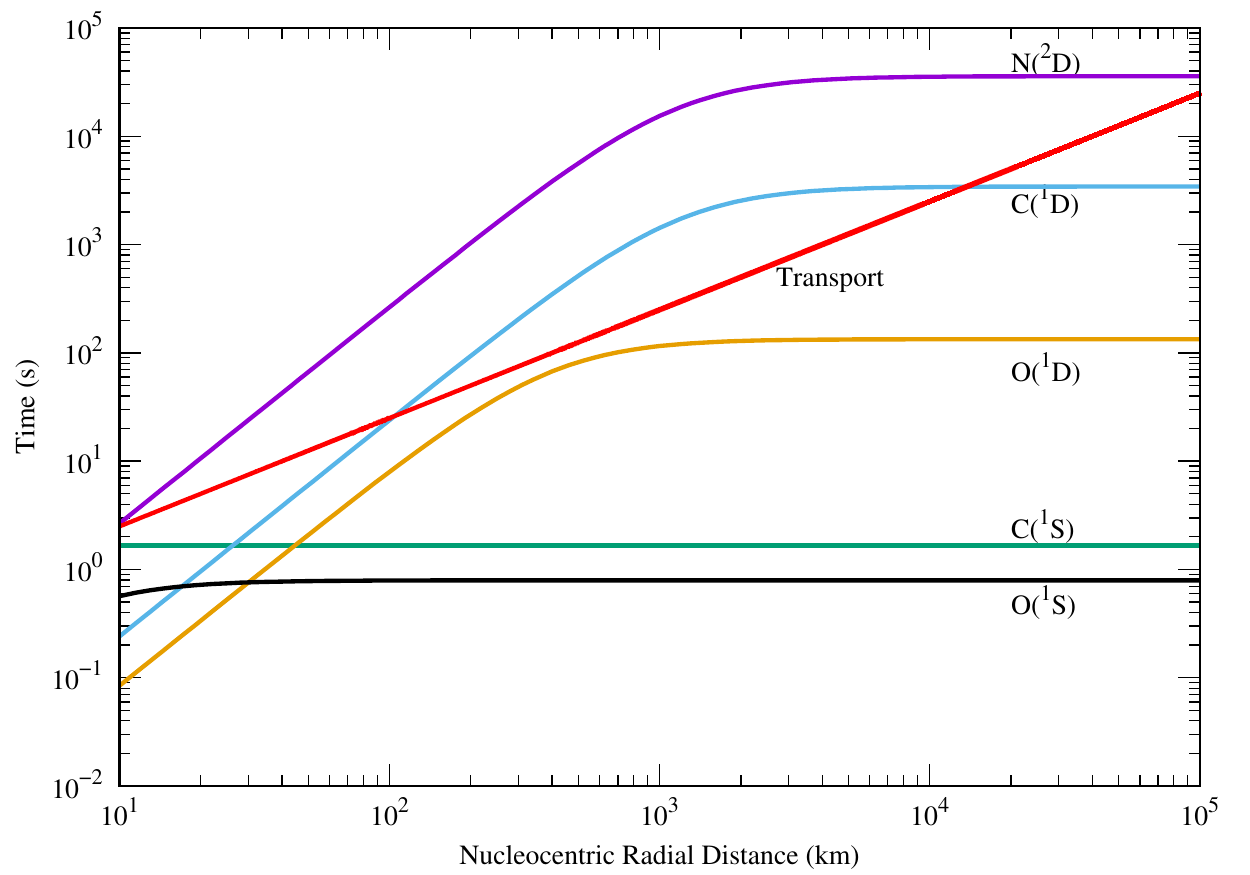}}
	\caption{Calculated chemical lifetime 
	profiles of N($^2$D), C($^1$D), O($^1$D), C($^1$S) and 
	O($^1$S) due to both collisional quenching and 
	radiative decay in C/2016 R2.  {For comparison, the timescale 
	for 
	transport of metastable species}
	(red line) is calculated for 2 km s$^{-1}$ radial 
	velocity. The input conditions are the same 
	as in Fig.~\ref{fig:pr_rate_o1sd}.}
	\label{fig:lifetime}%
\end{figure}

 {We have calculated the timescale profiles for
 the loss of excited states due to both collisional quenching 
 and radiative decay (chemical loss), which is reciprocal to the total loss 
 frequency
 profiles as 
 determined in Figs.~\ref{fig:ls_rate_o1sd}, 
 \ref{fig:ls_rate_c1n2d}, and \ref{fig:ls_rate_c1s}.  The timescale for 
 transport 
 (advection)
for the excited species is determined as 
t$_{adv}$(r) $\sim$ r $\times$ [2v$_j$(r)]$^{-1}$, where v$_j$ is the mean
radial velocity of the $jth$ excited species as determined from the observations
(see Table~\ref{tab:linewidth}).  These modelled timescale profiles
for chemical loss and transport are 
plotted in 
Fig.~\ref{fig:lifetime}.} It can be noticed in this 
figure that the O($^1$S) and C($^1$S) chemical lifetimes are smaller 
than  {the time required for transport} by more than an order 
of magnitude for radial distances larger than 100 km. The  {timescales} 
of O($^1$D)  {due to chemical loss and transport} are equal at the
radial distance of 300 km, whereas this equality occurs at 
a radial distances of  {100} and $\sim$2 $\times$ 10$^3$ 
km  for C($^1$D). The N($^2$D)  {timescale for transport} is 
smaller than the chemical lifetime in the entire coma.

\begin{figure}
	\centering
	\resizebox{\hsize}{!}{	
	\includegraphics{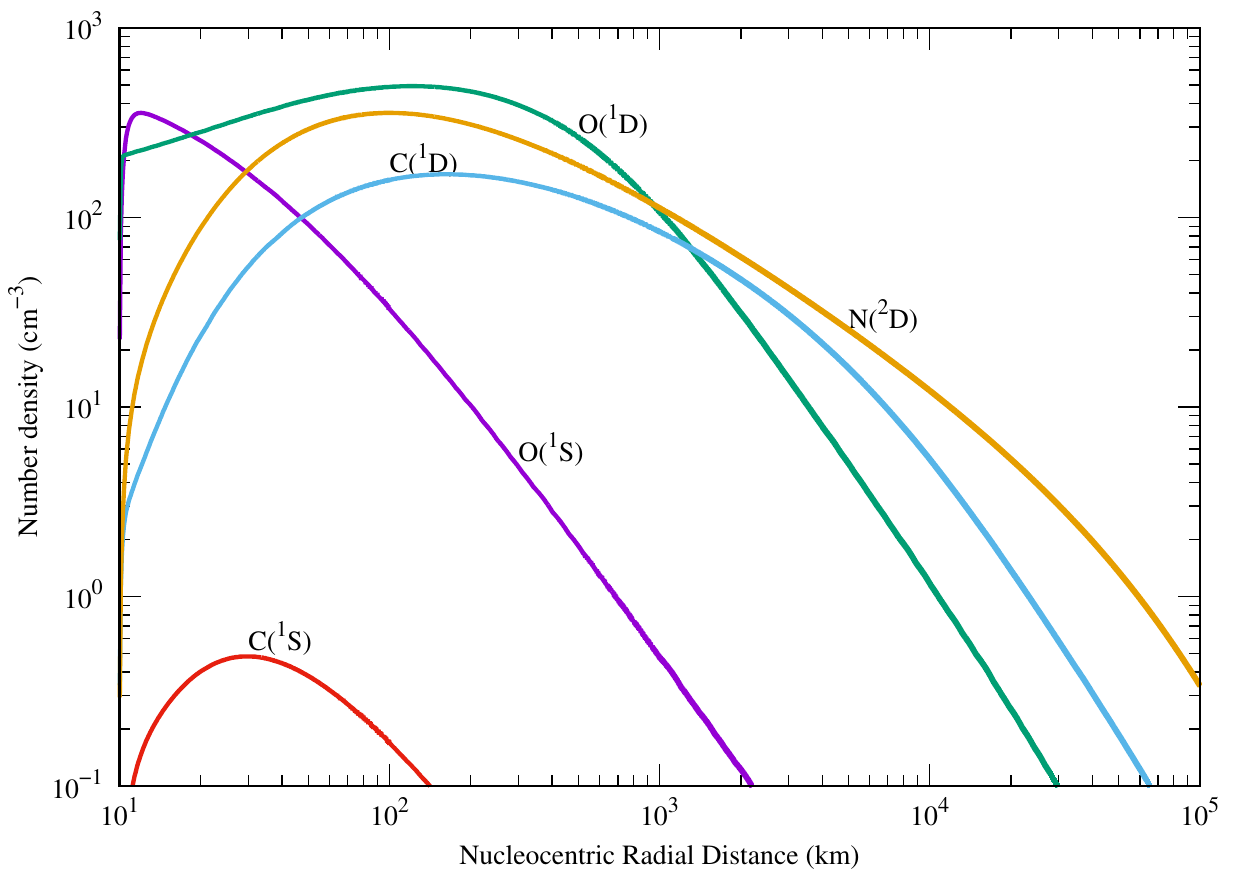}}
	\caption{Calculated number density profiles 
	of O($^1$S), O($^1$D), C($^1$D), N($^1$D), C($^1$S), by 
	incorporating  various photon and electron impact 
	dissociative excitation reactions of the major cometary 
	volatiles in C/2016 R2. The input conditions 
	are the same as in Fig.~\ref{fig:pr_rate_o1sd}.}
	\label{fig:nub_den_OCN}%
\end{figure}

After incorporating the previously mentioned  production 
and loss mechanisms, we have solved the one-dimensional 
continuity equation  {(Eq.~\ref{eq:conti})},
to calculate the radial density 
profiles of O($^1$S), O($^1$D), C($^1$D),  N($^2$D), and 
C($^1$S). 
They are plotted in Fig.~\ref{fig:nub_den_OCN}. 
The peaks of O($^1$S) and  {C($^1$S)} number densities occur
close to the surface of the  nucleus at radial distances below 
20 km. The calculated N($^2$D), C($^1$D) and O($^1$D)
 density  profiles have broad  peaks  
at around  100 km radial distance, due 
to their long radiative lifetimes. 

\begin{table*}[htb]
	\centering
\caption{ {Summary of the different case studies for calculating
       the  [OI] G/R and the [CI]/[NI]  emission ratios.}}	
   \resizebox{\hsize}{!}{%
\begin{tabular}{lll}
\toprule
Case& Description\\
\midrule
Case-A& By using  {the atomic and molecular parameters of various 
oxygen-bearing 
species} as described in 
 {Section~\ref{sec:atom_prop}}.\\
Case-B& For 30\% of O$_2$ relative abundance with respect to CO production 
rate, instead of 1\% of O$_2$ in the baseline parameters.\\
Case-C& The CO$_2$ photodissociation cross section producing O($^1$D) is 
increased by a factor of 3.\\
Case-D& By using  {the atomic and molecular parameters of various carbon 
and 
nitrogen-bearing 
	species} as described in  Section~\ref{sec:atom_prop}.\\
Case-E& The CO photodissociation cross section producing C($^1$D) is decreased
        by a factor of 4.\\
Case-F& The photodissociative excitation cross section of N$_2$ producing   
        N($^2$D) is increased by a factor of 3. \\
\bottomrule
\end{tabular}}
\label{tab:cases}
\end{table*}

\begin{figure}
	\centering
	\resizebox{\hsize}{!}{
	\includegraphics{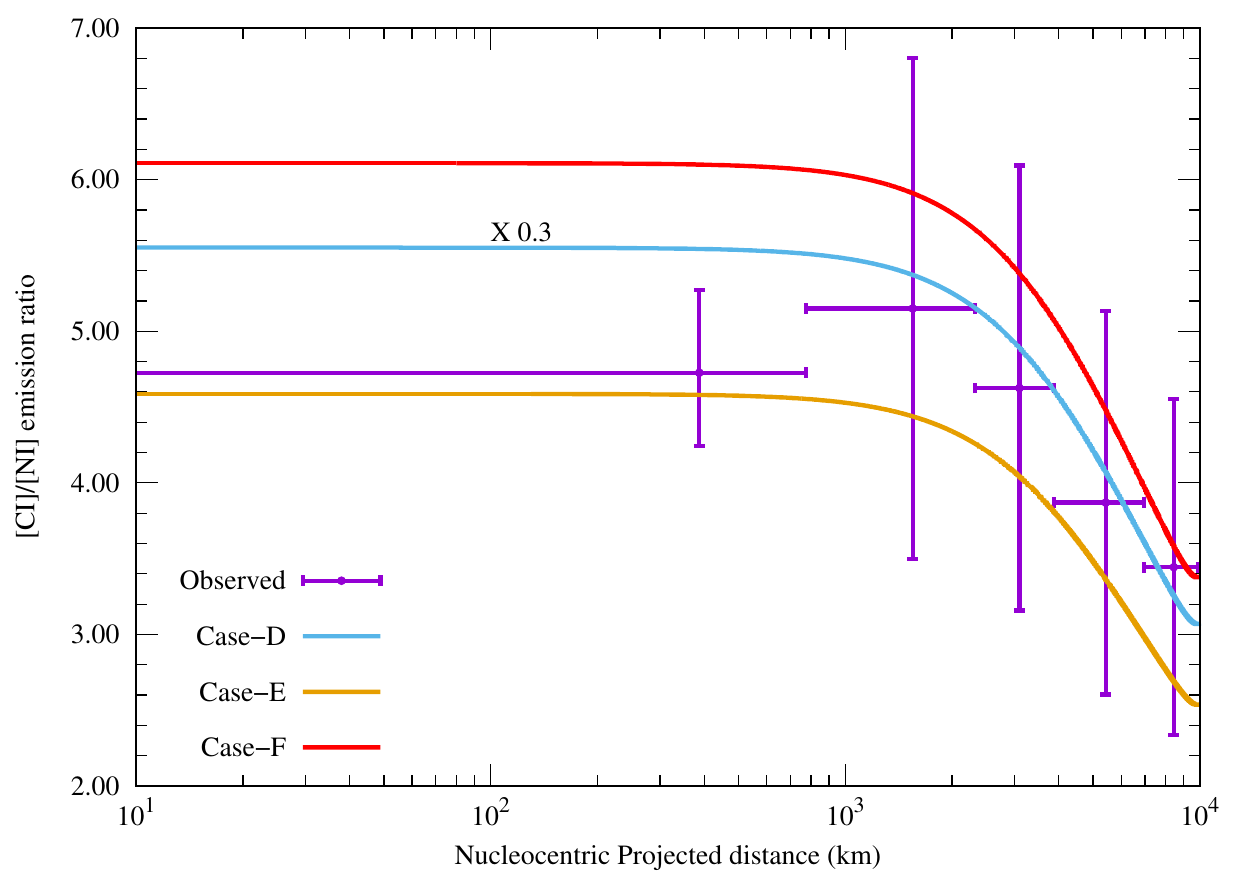}}
	\resizebox{\hsize}{!}{
	\includegraphics{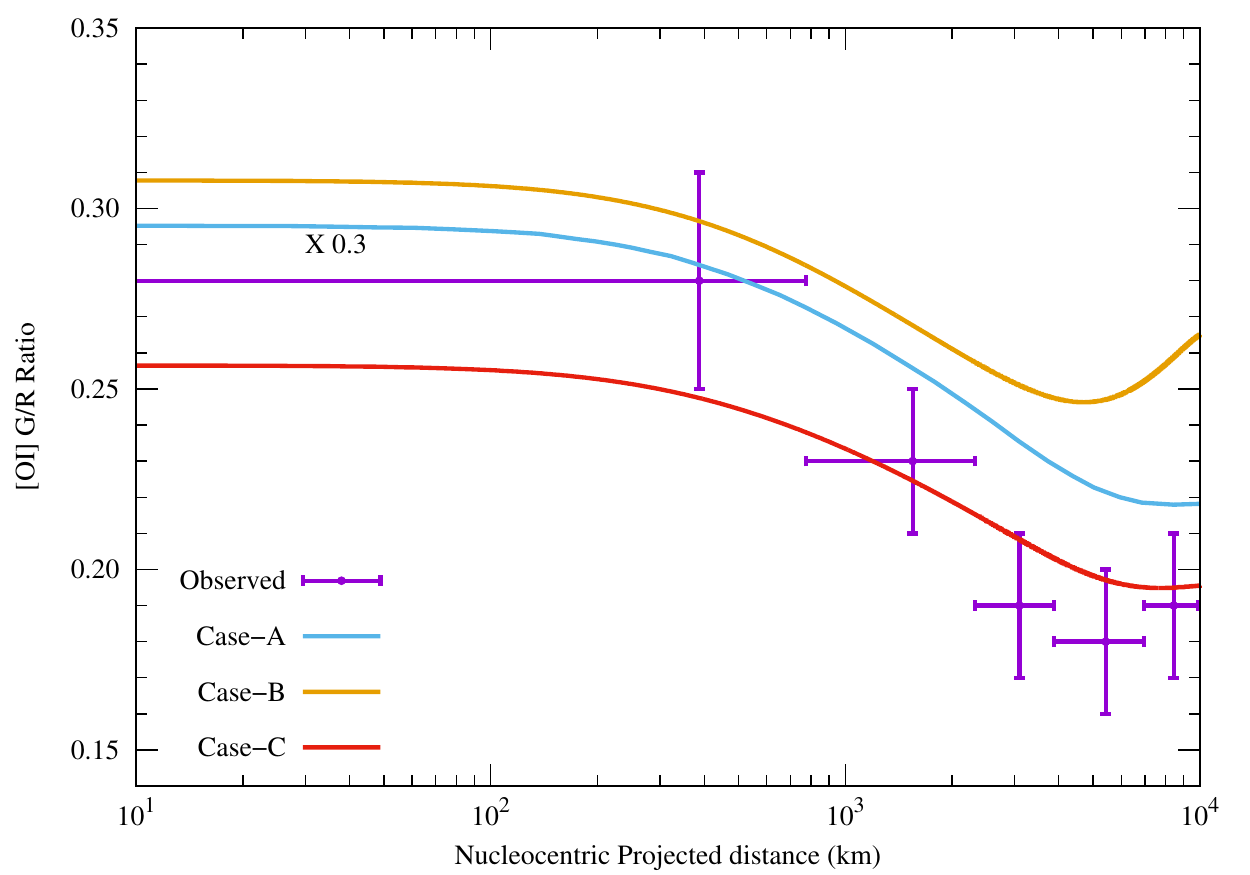}}
	\caption{Comparison between modelled and observed 
	 [CI] to [NI] (top panel) and [OI] G/R (bottom panel)
	emission ratios in C/2016 R2. The calculated emission 
	ratios for the input conditions are explained in 
	Fig.~\ref{fig:pr_rate_o1sd}.  {The calculated 
	emission ratios  {for Case-A and Case-D are plotted 
	after multiplication by a factor 0.3.} 
    See the main text and Table~\ref{tab:cases} 
    for more details about the conditions of 
    Case-A to Case-F.}} 
	\label{fig:gr_cn_ratio}%
\end{figure}

 {In the bottom panel of Fig.~\ref{fig:gr_cn_ratio}, 
we compare the modelled
[OI] green to red-doublet ratio with the observations.
Using the baseline parameters as discussed in Section~\ref{sec:model_ips},
we find that our modelled values are higher by a factor 3 when compared to 
observation (hereafter Case-A).  If we consider an O$_2$ 
abundance of 30\% with respect to the CO production rate (hereafter Case-B), 
instead of our
assumed 1\% in the baseline, 
our calculated [OI] G/R ratio is consistent with the observed emission 
ratio for radial distances below 1000 km. Above this radial 
distance the modelled ratio is  
about 40\% higher than the observations. In this case, 
photodissociation of CO$_2$ and O$_2$ are the major sources of  
O($^1$S) and O($^1$D), respectively. Due to the uncertainty in the CO$_2$ 
photodissociative excitation cross section producing 
O($^1$D), as discussed in the next section, we 
have increased the photodissociation rate by a factor of 
about 3 in our model calculations (hereafter Case-C). For this case, the 
modelled [OI] green to red-doublet emission ratio is in agreement with the 
observations.}

 {Similarly, by using the baseline parameters as discussed in 
Section~\ref{sec:model_ips}, the modelled [CI]/[NI] 
emission ratio profiles are found to be higher by a factor 3 than the 
observations (hereafter Case-D, and see top panel of 
Fig.~\ref{fig:gr_cn_ratio}). Due to the uncertainty in the photodissociative 
excitation cross section, we have decreased the photodissociation 
rate of CO producing C($^1$D) by a factor of 4 (hereafter 
Case-E) in the model calculations. With these new dissociation 
rates, we find that  the calculated emission ratios 
are  {in agreement with} the observations.  When we increase the 
N$_2$ dissociative excitation rate producing N($^2$D) by a 
factor of about 3 (hereafter Case-F), we find that the 
modelled [CI]/[NI] emission ratio is consistent with the upper limit 
of the observations. The conditions for different case studies are 
summarized in 
Table~\ref{tab:cases}.}

\begin{figure}
	\centering
	\resizebox{\hsize}{!}{	
	\includegraphics{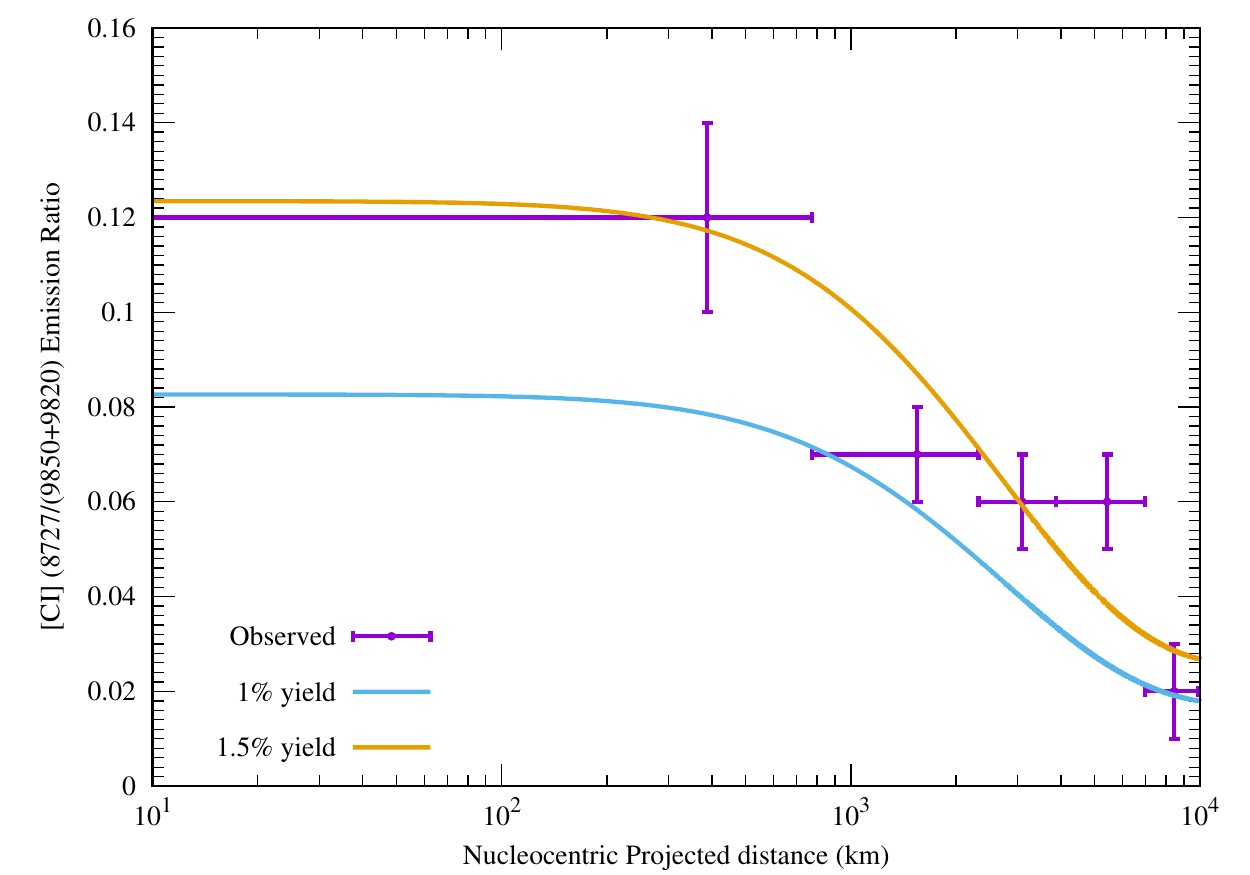}}
	\caption{Comparison between modelled and observed 
	[CI] 8727 \AA/(9850 \AA + 9824 \AA) emission 
	intensity ratio profiles in C/2016 R2. Solid curves
	represent the modelled emission ratios for two 
	different photodissociation yields of C($^1$S).} 
	\label{fig:[CI]_ratio}%
\end{figure}

If we assume that CO is the only parent source of the atomic 
carbon emissions, then the observed [CI] ratio should be 
equal to the ratio of the average branching fractions of CO 
producing C($^1$S) and C($^1$D). As described in the 
earlier section, we have modelled the C($^1$S) density and 
calculated the 8727 \AA\ emission intensity by assuming 
that 0.5\% of the total CO absorption cross section leads  
to C($^1$S) formation. Our modelled [CI] emission ratios 
for different yields along with the observation are plotted 
in Fig.~\ref{fig:[CI]_ratio} as a  function of the 
nucleocentric projected distance. We find that 
the calculated [CI] 
ratio profiles are  {consistent with} the observations when we vary 
the CO photodissociation yield between 1 and 1.5\%.

\section{Discussion}
 \label{sec:discuss}

The photochemical processes such as dissociative excitation 
of cometary coma species, collisional chemistry and radial 
transport of metastable species determine the observed 
emission intensity ratios. When we use our baseline input 
parameters, as explained in Section~\ref{sec:model_ips}, 
the calculated [OI] G/R  and [CI]/[NI] ratios 
are higher by a factor of about 3 compared to 
the observations. In order to study the 
discrepancies between the modelled and the observed 
emission ratios, we examined the effects of  
uncertainties associated with the input parameters, which 
are discussed below.

\subsection{Effect of the neutral species abundances}

In a water-dominated coma, H$_2$O photodissociation is 
the major source of O($^1$D). But the observations of 
\cite{Biver18} showed that 
H$_2$O abundance with respect to CO can not be more 
than 10\%. The recent 
\cite{McKay19} observations suggest much lower water
abundance (0.3\% with respect to CO), which is also 
observed by \cite{Opitom19}.
Even by assuming an H$_2$O abundance of 10\% relative to CO,
we find that H$_2$O is playing no role in determining the [OI] G/R ratio. 
This calculation suggests that CO$_2$ mainly governs  {both the green
and red-doublet} atomic 
oxygen emission intensities and the contribution from other excitation
processes is negligible. The derivation of water
production rates based on the observed oxygen emission lines
would of course lead to an over estimation of H$_2$O abundance 
in this comet. 

\cite{Biver18} observed several other O-bearing species 
such as CH$_3$OH, H$_2$CO, HNCO, NH$_2$CHO, CH$_3$CHO, 
HCOOH, SO, and SO$_2$ in this comet.  {The branching ratios
as determined by \cite{Huebner92} suggest that} these species do 
not have direct production channels for O($^1$S) and 
O($^1$D). Moreover, these species are 
observed with very low abundance, more than two orders of 
magnitude smaller than CO  {\citep{Biver18}}.
 {Hence, the contribution of
these species producing [OI] emission lines can be neglected.}

The modelling work of 
\cite{Cessateur16a} has suggested that O$_2$ can play a 
significant role in determining the [OI] G/R ratio. 
By incorporating a small O$_2$ abundance, of 1\% of the 
CO production rate, we find that the contribution of O$_2$ 
dissociative excitation is very small (more than two orders 
of magnitude to the total) in producing O($^1$S) whereas it 
is the third most important source of O($^1$D) for radial 
distances above 30 km (see Fig~\ref{fig:pr_rate_o1sd}). 
By increasing the O$_2$ abundance to 4\% of CO,
we find that about 30\% of O($^1$D) is produced from the
photodissociation of O$_2$ and 50\% of the total is coming 
from CO$_2$.  In this case, O($^1$S) is mainly ($>$ 80\%) 
produced from the photodissociation of CO$_2$ and the role 
of O$_2$ is small ($<$ 1\%). This calculation suggests 
that O$_2$ is a potential source of O($^1$D), after 
CO$_2$, when its abundance is substantial ($>$5\%).

The in-situ measurements of the Rosetta mass 
spectrometer  have shown that  {O$_2$ is present 
in the coma of comet 67P with a mean abundance of about 2\% 
\citep{Altwegg19}. During the Rosetta mission, \cite{Bieler15} have
also shown that the O$_2$ 
relative abundance with respect to H$_2$O can be as high as 15\%.}
 It should be noted that the 
mass spectrometer abundances are 
based on the observations made at the spacecraft 
position. However, the Alice UV spectrometer on-board Rosetta 
can observe  {spatial distribution of neutrals in the entire} cometary 
coma based on the emission intensities \citep{Feldman18}. 
From Alice  {limb view observations}, \cite{Keeney17} found that the
O$_2$/H$_2$O relative abundance ratio in 67P coma 
varies between 11–68\%, with a mean value of 25\%. These 
observations suggest  that O$_2$ could be present in large 
amount in cometary comae. We then have increased the O$_2$ relative 
abundance to 30\% of CO to verify its impact on the 
[OI] G/R abundance ratio.  By considering such a large O$_2$ 
abundances, we have shown that the model 
calculated [OI] G/R ratio is close to the observation
(see bottom panel of Fig.~\ref{fig:gr_cn_ratio}) for radial 
distances below 10$^3$ km. However, the calculated 
emission ratios are higher by about 40\% compared to the 
observation in the non-collisional zone,  i.e., for radial 
distances larger than 10$^3$ km (also see 
Fig.~\ref{fig:ls_rate_o1sd}). This calculation suggests 
that there is a possibility of having a significant 
amount of molecular oxygen (about 30\% relative to CO) 
in this comet.

Similarly, we have studied the discrepancy between the 
modelled and observed [CI]/[NI] emission ratio by varying 
the neutral abundances. 
We have considered CO$_2$ and CO as the primary molecules to 
produce C($^1$D) and  {CN and} N$_2$ for N($^2$D). The role of 
CO$_2$ in producing C($^1$D) is negligible compared to that 
of CO. This is  {mainly} due to the significant difference  between the 
photodissociative 
excitation cross sections of CO and CO$_2$  {producing atomic carbon
 \citep[about two orders of 
magnitude, see][]{Huebner92}.} C($^1$D) can 
be produced 
directly from the dissociation of CO whereas it is a 
two-step process for CO$_2$. By increasing the CO$_2$ 
abundance equal to that of CO, we find that  CO is still 
the dominant source of C($^1$D), which suggests that [CI] 
emission lines are governed only by the CO photodissociation
and the role of CO$_2$ in producing carbon emission lines can be
neglected. 

In order to match the observed [CI]/[NI] emission ratio, 
the N$_2$/CO volume mixing ratio should be altered by a 
factor  {of 3} either by decreasing the CO or by increasing the 
N$_2$ production rates. Since the comet is moving towards 
the Sun, we do not expect a decrease in the CO sublimation 
rate during the observation period. However, large 
 N$_2$ abundances ($>$10\%) are not expected either, unless 
there is a cometary outburst during the period of 
observation. Moreover, different observations have 
confirmed that the N$_2$ abundance in this comet is 
about 7\% relative to CO \citep{Cochran18, Biver18, 
Opitom19}. Hence, we do not expect this comet to have a 
large enough N$_2$ abundance (>10\%)  to produce the observed 
[CI]/[NI] emission ratio.
 We have also investigated  other 
N-bearing species which can produce N($^2$D) with a
significant amount. \cite{Singh91} proposed that the formation of 
metastable nitrogen atoms in $^2$D state is possible due to the
photodissociation of CN.  {But our modelled production rate profiles of 
N($^2$D) 
show that the contribution from CN is negligible compared to the 
photodissociation 
of N$_2$ (see Fig.~\ref{fig:pr_rate_c1dn2d}). Moreover,} \citet{Opitom19} 
estimated the CN 
production rate as 3 $\times$ 10$^{24}$ s$^{-1}$, which is 
five orders of magnitude smaller than the CO gas production 
rate and cannot be the dominant source of
[CI] and [NI] emission lines. 
 Similarly, the observations of various 
N-bearing species viz., HCN, CH$_3$CN, HC$_3$N, HNCO, 
NH$_2$CHO by \cite{Biver18} and \cite{McKay19} 
show that their relative abundances in the coma are very 
small (<1\%) compared to N$_2$,  {while the \cite{Huebner92} 
compilation of photodissociation cross sections does not suggest 
direct dissociative excitation channels to produce 
N($^2$D)}. Hence, we conclude that the atomic 
nitrogen emission is only governed by the photodissociation 
of N$_2$ in C/2016 R2.

The \cite{Opitom19} observations 
clearly show that C$_2$ and C$_3$ molecules are only present 
in negligible amount in this comet. 
 {Unlike the major volatiles, C$_2$ and C$_3$ are not directly
produced in the coma via sublimation of ices in the nucleus.
The formation and destruction of these species is involved with 
a complex chemistry. By considering the long-chain carbon-bearing 
species in the coma, \cite{Alexander15} has studied the density distribution
of  C$_2$ and C$_3$ in the comae of different comets. Based on these 
calculations, we estimated that the production rate of C($^1$D) via 
dissociation of C$_2$ and C$_3$ is smaller than the photodissociation rate of CO
 by more than three orders of magnitudes. However, 
the calculation of C$_2$ and C$_3$ number density profiles using the 
complex chemical network as described by \cite{Alexander15} is beyond the 
 scope of the  present work.}

\subsection{Effect of the cross sections}
The uncertainty in the photodissociation cross section of 
CO$_2$ producing O($^1$D) could be the reason of the 
difference between the modelled and the observed 
[OI] G/R ratio. \cite{Lawrence72a} experimentally 
determined the photodissociation yield of CO$_2$  
producing O($^1$S) as high as 1 in the wavelength range 
1000--1150 \AA. We have considered this measured yield to 
calculate the CO$_2$ dissociation rate to produce O($^1$S). 
The cross section for O($^1$D) formation via CO$_2$ 
dissociative excitation is not reported in the literature, 
as noticed by \cite{Huestis06}. \cite{Huebner92} 
constructed these photodissociation cross sections of CO$_2$ assuming a dissociation branching 
ratio close to 1 in the wavelength region 1000--1665 
\AA. This assumption contradicts the \cite{Lawrence72a} 
experimentally determined branching ratio. Both O($^1$D) 
and O($^1$S) can not be produced with 100\% yield in the 
same wavelength region. \cite{Slanger77} also measured the 
quantum yield of O($^1$S) for photodissociation of CO$_2$ 
and found a sudden dip in the yield around 1089 \AA. 
Hence, there is an uncertainty associated with the
\cite{Huebner92} recommended cross sections which can 
impact our calculated [OI] G/R emission ratio. There 
are several recent experimental developments in detecting 
O($^1$D) via the photodissociation of CO$_2$ 
\citep{Sutradhar17, Lu15, 
Song14, Schmidt13, Gao19, Pan11}. But these measurements 
are limited to a small energy range ($<$1 eV) and do not provide 
the O($^1$D) yield over a wide wavelength range 
(from dissociation threshold to 1000 \AA) which is essential 
to determine the average dissociation ratio of O($^1$S)/O($^1$D)
in the photodissociation of CO$_2$.  

As discussed in the earlier section, the photodissociation 
of CO$_2$ is the major source of O($^1$S) in the coma. This 
is mainly due to the larger 
 {photodissociation rate of CO$_2$ than CO and H$_2$O
\citep[a 
factor around 20 higher compared to other 
O-bearing species, see Table~\ref{tab:phot_freq}, 
Fig.~\ref{fig:pr_rate_o1sd}, and][] 
{Huebner92}.}
 Similarly, CO$_2$ can also
produce O($^1$D) by a factor of 3 more efficiently than any other 
O-bearing species. Hence, if CO$_2$ is the major production 
source of oxygen lines, then the observed [OI] 
G/R ratio in the non-collisional region (i.e. above 10$^3$ 
km, see Fig.~\ref{fig:ls_rate_o1sd}) should match  the 
ratio of CO$_2$ dissociation rates producing O($^1$S) and 
O($^1$D). 

Our calculations suggest that the photodissociation of CO is the 
second most important source of [OI] emission lines (see 
Fig.~\ref{fig:pr_rate_o1sd}). Since 
most of the O($^1$S) and O($^1$D) are produced from CO$_2$, the
uncertainties associated with the CO dissociation cross sections
 can not play a significant role in determining
the [OI] G/R ratio. Moreover, the respective
dissociation threshold energies of CO, producing
O($^1$D) and O($^1$S) states are 14.35 and 16.58 eV, which are more than the 
ionization threshold i.e., 14 eV. Due to the proximity of the threshold 
energies, most of the UV
radiation absorbed by CO turn into ion rather than producing O($^1$S) and
O($^1$D). Hence CO is not an efficient source of [OI] emission lines,
when compared to CO$_2$, to determine the [OI] G/R ratio.

In our earlier work, we have theoretically determined [OI] 
G/R ratios of 0.04, 0.6, 1, and 0.04  for the photodissociation 
of H$_2$O, CO$_2$, CO, and O$_2$, respectively \citep{Bhardwaj12, 
Raghuram13, Raghuram16}. If CO$_2$ is the only source of 
these emission lines,  {by using the photodissociation
cross sections
from the baseline model \cite[also see][]{Bhardwaj12},  our 
model suggests that the} [OI] G/R ratio 
should be around 0.6 for the non-collisional region. But the 
observed ratio  {in this comet} is only about 0.2 between radial 
distances of 2 
$\times$ 10$^3$ 
and 1 $\times$ 10$^4$ km, where the radiative decay is the 
only loss process  {(see~Fig.~\ref{fig:gr_cn_ratio})}. We will show in 
the next 
section that 
the  collisional quenching can alter the observed [OI] G/R 
ratio only for radial distances less than 2 $\times$ 10$^3$ 
km  (see~Fig. \ref{fig:ls_rate_o1sd}). Since CO$_2$ is the 
major source of [OI] emissions and there is a large 
uncertainty associated with the CO$_2$  photodissociative 
cross section in producing O($^1$D),  we have  {increased} the 
photodissociation rate by a factor of 3 (see  {Case-C} in 
Fig.~\ref{fig:gr_cn_ratio}). For this value of the 
photodissociation rate, the observed and modelled 
[OI] G/R ratios are  {consistent}.  {
Hence, based on the comparison between the modelled and observed 
[OI] G/R ratio profiles, we suggest to increase 
the \cite{Huebner92} recommended photodissociation cross section 
of CO$_2$ producing O($^1$D) by a factor of 3.}

In our modelling calculations on different comets at 
different heliocentric distances, we have 
shown that H$_2$O is the major source of O($^1$D) 
\citep{Bhardwaj12, Raghuram13, Raghuram14, Decock15, 
Raghuram16}. Even though the photodissociation rate of H$_2$O 
producing O($^1$D) is comparable (a factor of 1.5 higher) 
to that of CO$_2$ \citep[see,][]{Huebner92}, CO$_2$ became the 
prominent source of green 
and red-doublet emission lines  due to the very low 
water abundance in this comet. Hence, this new value  
of CO$_2$ dissociation rate producing O($^1$D) does not
contradict the conclusions of our earlier calculations done in the 
water-dominated cometary comae. 

Similarly, the discrepancy between the modelled and the 
observed [CI]/[NI] emission ratios could also be due to 
the uncertainties in the photodissociation cross 
sections of CO and N$_2$. \cite{Shi18} recently determined the branching 
ratio for CO producing C($^1$D) in the wavelength range 
905--925 \AA. They found that the pre-dissociation of CO 
leads to  about 20\% formation of C($^1$D) and the rest 
into carbon and oxygen ground and $^1$D states, 
respectively. \cite{Huebner92} do not account for 
pre-dissociation channels in the formation of C($^1$D). The 
cross section is constructed based on the assumptions of 
\cite{Mcelroy71}.  By decreasing the CO photodissociation 
rate by a factor  {of 4}, we find that the modelled [CI]/[NI] 
emission ratio is  {consistent with}  the observation (see 
 {Case-E} 
in Fig.~\ref{fig:gr_cn_ratio}). This calculation shows 
that the uncertainty associated with the \cite{Huebner92} 
CO photodissociation cross section could be a reason for 
the discrepancy between the modelled and the observed 
ratios. So we suggest that the photodissociation cross 
section of CO producing C($^1$D) should be revised based on 
the recently determined branching ratios. 

We have taken a mean average branching ratio of N($^2$D) of 
0.32 following the approach of \cite{Fox93} calculations on 
Mars and Venus atmospheres. \cite{Song16} measured 
branching ratios for the photodissociation of N$_2$ 
producing N($^2$D) in the wavelength region 816.3--963.9 
\AA\ as high as 1.  {By increasing the N$_2$ 
photodissociation rate by a factor of 3, we find that  the 
modelled [CI]/[NI] emission ratio is  {consistent with}
the upper limit of the observations (see  {Case-F} in 
Fig.~\ref{fig:gr_cn_ratio})}. These calculations suggest 
that the uncertainties associated with the photodissociation 
cross sections of CO and N$_2$ are the main reason for the 
discrepancy between the modelled and the observed [CI]/[NI] 
emission ratios. 

 {Besides the cross sections of neutral species, the modelled 
photodissociation 
frequencies also depend on the solar activity. Considering the solar activity, 
\cite{Huebner92} have shown that the calculated photodissociation 		
frequencies of neutrals can vary by a factor of about 2 from solar minimum 
to maximum. Since the observations of C/2016 R2 are all done in 
February 2018,  {which is during a minimum range of time and} solar 
minimum 
condition, we do not expect
the solar activity to significantly influence the modelled dissociation 
frequencies and subsequently the calculated emission ratios.}

\subsection{Effect of collisional quenching rates}
The collisional quenching of atomic oxygen can alter the 
observed emission ratio for radial distances smaller than 2 
$\times$ 10$^3$ km (see Fig.~\ref{fig:lifetime}).  {Owing to
a short radiative lifetime  \citep[0.75 s,][]{Wiese09}, 
the collisional quenching of O($^1$S) is not significant in the entire coma. 
But for O($^1$D), the quenching is an important loss process up to radial 
distances of 2 $\times$ 
10$^3$ km (see Fig.~\ref{fig:ls_rate_o1sd}).}
By assuming a collisional quenching 
source of O($^1$D), which is  an order of magnitude higher 
than that of CO$_2$,  we could alter the G/R ratio only up to
radial distances $\simeq$10$^3$ km by a factor of 2. Hence, the observed
[OI] G/R ratio, beyond 10$^3$ km radial distances, is purely 
a function of dissociative excitation of CO$_2$.   

The collisional quenching is significant for both C($^1$D)
and N($^2$D) due to their radiative lifetimes of about  {3500 
s and}  {10} hrs, respectively, up to radial distances of 2 
$\times$ 10$^3$ km (see Fig.~\ref{fig:ls_rate_c1n2d}). But we 
could not find a significant change in the calculated 
[CI]/[NI] emission ratios by varying the quenching rates. 
Our calculated lifetime profiles of these excited species 
show  that the transport of N($^2$D) is a more important loss 
process than the collisional quenching in the coma (see 
Fig.~\ref{fig:lifetime}). In this figure, it can be noticed 
that the C($^1$D) collisional quenching can affect the 
modelled [CI]/[NI] emission ratio only for radial distances 
below 50 km.  {By assuming an unknown collisional reaction in the coma 
with a quenching rate of an order magnitude higher than that of 
CO, we  find the modelled [CI]/[NI] emission ratio is not changing 
for radial distances beyond 4 $\times$ 10$^3$ km.}

The collisional quenching is a strong barrier inhibition in measuring
the photodissociation crosssection of species producing
metastable states. Due to the long lifetimes and also the
excess velocities acquired during dissociation, most of the
metastable species drift away from the  spectrometer field of 
view and also collide with the walls of the chamber which makes 
very difficult to measure such cross sections in 
the laboratory \citep{Kanik03}. Owing to natural vacuum
conditions, the metastable species produced via cometary
species can survive longer times and will be transported to
large radial distances from the origin of production before 
they de-excite to ground state via radiative decay. Hence in 
the absence of collisions, the observed cometary forbidden
emissions lines can be used to study the dissociation 
properties of the respective parent species which makes comets 
unique natural laboratories.

\subsection{Effect of transport}
When the mean lifetime of a species is more controlled by 
collisions or radiative decay than radial transport, 
we can consider that the species is under photochemical 
equilibrium (PCE). The calculated lifetime profiles presented in 
Fig.~\ref{fig:lifetime} show that O($^1$S) and C($^1$S)
are in PCE in the entire cometary coma. The transport 
of C($^1$D) and O($^1$D) is important for radial distances 
above  {50 and 100 km}, respectively. Due to the large 
radiative lifetime of N($^2$D)  {(about 10 hrs)}, the 
transport plays a significant role in determining its  
number density in the coma. These calculations suggest that 
the role of transport in determining the [OI] G/R ratio is 
insignificant whereas it is important for the [CI]/[NI] 
emission ratio.

 {\cite{Huebner92} have determined the mean excess energies released 
in the photodissociation of N$_2$ and CO as 3.38 and 2.29 eV, respectively.}
 {Using these calculated excess energies,} we estimated the mean 
ejection velocities of 
N($^2$D) and C($^1$D) as 4.77 and 4.59 km/s, respectively, { 
\citep[cf.][]{Wu93}.} 
By using these theoretically estimated  mean velocities in the model, 
we found our modelled [CI]/[NI] emission ratio 
is higher by a factor of 2. If we assume that our estimated mean excess 
velocities of C($^1$D) and N($^2$D) are realistic, then a 
factor of 2 in CO photodissociation rate 
will be sufficient to explain the observed [CI]/[NI] 
ratio. But our observed line widths suggest that 
C($^1$D) and N($^2$D) did not acquire velocities as 
calculated by \cite{Huebner92}.

Our theoretically estimated N($^2$D) and C($^1$D) excess 
velocities are based on the calculation of 
\cite{Huebner92}. These values are higher than our 
mean velocities  derived  {from the observations} 
by about a factor of 2  {(see Table~\ref{tab:linewidth})}. However, it 
should be 
noticed that \cite{Huebner92}  {have used a reference
solar flux for solar minimum condition to 
determine the excess energy released in the photodissociation. The solar 
flux as observed by the comet determines the mean excess energy released
in the photodissociation and subsequently the widths of the spectral lines}.
 The 
determination of excess energies released in the 
photochemical process requires the knowledge of the 
solar radiation flux interacting with the coma and also the absolute 
branching fractions over a wide wavelength range, which 
are the strong constraints to compute the mean excess 
velocities.  

The role of transport on O($^1$D) and O($^1$S) is  
not significant in determining the [OI] G/R ratio. This 
can be understood from the  {calculations of the timescales
for chemical lifetime and transport presented 
in}   Fig.~\ref{fig:lifetime}. We have calculated density 
profiles of all the metastable species using the derived 
excess velocities from the observed line widths.
The \cite{Huebner92} theoretical calculations at 1 au  
suggest that excess energies of about 4 eV will be released 
in the photodissociation of CO$_2$ producing O($^1$D). 
Around 65\% of excess energy released in the dissociation
would produce atomic oxygen with an excess velocity of 
about 5 km/s. By increasing the radial 
velocity of the metastable species  to about 5 km/s, our 
calculated [OI] G/R ratio is higher by a factor of 5 
compared to the observations. The role of transport in
determining the O($^1$S) density is negligible due to its small 
lifetime of about 1 s, whereas it can affect O($^1$D) up to 
radial distances of 10$^4$ km. Due to the lack of measured 
cross sections of O($^1$D) via CO$_2$ it is difficult to 
determine the excess energy released in the photodissociation. 

\subsection{Line widths} 
The observed FWHMs of the emission lines mainly depend on the 
mean  {kinetic} energy of the excited species which are
produced in the dissociative excitation. Our calculations 
suggest that the influence of collisional quenching on the 
line width is up to radial distances of 10$^3$ km 
(see Figs.  \ref{fig:ls_rate_o1sd} and
\ref{fig:pr_rate_c1dn2d}). In the observed spectra, we 
find that the measured line widths of the observed 
emissions are not varying significantly as a function 
of the nucleocentric distance. Moreover, the radiative 
efficiency of all the metastable states,  except 
for O($^1$S), is close to unity for radial distances above 
10$^3$ km,  where most of the emission intensity occurs. 
Since the emission intensity in the observed spectra 
is mostly determined by the radiative decay of the 
excited species, we can neglect the contribution of 
collisional broadening while converting the line width to the 
mean velocity of the excited species. 

The line widths reported in Table~\ref{tab:linewidth} 
show that the atomic oxygen green line is broader than 
the red-doublet lines, and this is also consistent with 
the observation made in other comets \citep{Cochran08, 
Decock13, Decock15}. This can be 
explained by the excess velocity of oxygen atom
acquired during the CO$_2$ photodissociation. The formation 
of O($^1$S) via CO$_2$ dissociative excitation occurs at 
shorter wavelengths compared to that of O($^1$D), which 
leads to more excess energy release and results in higher 
mean velocity. \cite{Raghuram13} calculated the 
mean excess energy profiles of O($^1$S) and O($^1$D) 
in comet Hale-Bopp and showed that the photodissociative 
excitation reactions produce O($^1$S) with  excess 
energies larger than for O($^1$D). Based on the measured line 
widths, \cite{Cochran08} suggested that the larger green 
line width could be due to the involvement of different 
energetic photons in dissociating a 
single parent species and/or dissociation of multiple species, 
producing O($^1$S) and O($^1$D) with different excess
energies. Our model calculations suggest that CO$_2$ mainly produces 
both green and red-doublet emission lines and the 
involvement of other species is small 
(see Fig.~\ref{fig:pr_rate_o1sd}). Hence, 
we argue that the observed green line width is larger than
the red-doublet widths mainly because  CO$_2$ is dissociated 
by high energetic solar photons which produce O($^1$S) with 
large excess velocities.

 Similarly, the width of [CI] 8727 \AA\ line is larger
 than that of 9850 \AA\ line and can also be due to the involvement 
 of high energetic photons in the photodissociation of CO. 
 The dissociation thresholds of CO producing C($^1$D) and
 C($^1$S) are 12 and 14 eV, respectively. Hence, 
 the CO dissociation by high energetic photons leads to more
 excess velocity for C($^1$S) than for C($^1$D). 
 Moreover, the production rate of atomic carbon via
 CO$_2$ dissociative excitation is very low compared to 
 that of CO \citep{Huebner92}. Hence, the role of CO$_2$ 
 in the C($^1$S) formation  can be neglected.    
 As shown in Fig.~\ref{fig:eleves_diag}, the 
 radiative decay of C($^1$P) can also populate 
 C($^1$D), which additionally contribute to the
 emission intensities at wavelengths 9824 and 9850 \AA. 
 Since this formation channel occurs at energies higher 
 than 18 eV, where most of the CO turns into ion, the 
 contribution from radiative decay of C($^1$P) can be 
 neglected in the formation of C($^1$S).

Our observed red-doublet emissions' line widths are 
similar, which suggests that the O($^1$D) is released 
with a mean excess velocity of 2.01 km/s from dissociative 
excitation (see Table~\ref{tab:linewidth}). The derived
line widths for the 5198 and 5200 \AA\ emissions, which
arise due to radiative decay of same excited state, i.e.,
N($^2$D) (see Fig.~\ref{fig:eleves_diag}), are 
 {compatible with the uncertainties in the observation.}

\subsection{On the doublet-emission ratios}
\label{sec:doub_rate}
The radiative decay of metastable states viz., C($^1$D),
O($^1$D), and N($^2$D) to the ground state produces doublet 
emissions due to splitting in the sub-levels of the electronic
states (see Fig.~\ref{fig:eleves_diag}). Hence, the 
observed intensity ratio of these doublet emission lines
should be equal to the ratio of the Einstein decay
probability coefficients of the corresponding transitions.
The [CI] 9824 \AA\ emission line is strongly contaminated 
by  {telluric absorption line} which
did not allow us to determine precisely the [CI] 9824/9850 emission
ratio. On the contrary the measured [OI] ratio over the full slit is
well determined and is  {in agreement with} the theoretical value
\citep{Wiese96}. Based on the theoretical calculations of 
\citet{Tachiev02}, \cite{Wiese07} recommend a transition 
probability ratio of the [NI] doublet equal to 2.6. But our
 observations show that this
value is 1.22 $\pm$ 0.09, which is nearly half the 
\cite{Wiese07} value. \cite{Sharpee05}
noticed that there is a significant difference between the
calculated and observed [NI] emission ratios. 
They measured the [NI] emission ratio in the night glow of the
terrestrial spectrum and found it has an average value
of 1.76. In the terrestrial atmosphere, strong 
collisions between N($^2$D) and other species can alter the 
statistical population of atomic nitrogen excited levels,
which subsequently affects the intensity ratio of the [NI]
emission lines. The \citet{Hernandez69} and \citet{Sivjee81}
measurements have  shown that the [NI] ratio varies between 
1.2 and 1.9. \citet{Sivjee81} calculated an emission ratio profile
by considering major collisional quenching reactions 
of N($^2$D). They found that [NI] ratio can be as low as
1.1. Our calculated lifetime profiles show that
throughout the coma, the N($^2$D) density is essentially 
determined by transport rather than quenching due to the poor 
collisional interactions with other cometary species (see
Fig.~\ref{fig:lifetime}). Hence, the role of collisional
quenching in the observed [NI] ratio can be neglected. Based
on the observations and modelling, we argue that the measured
[NI] emission ratio is mainly due to the intrinsic transition
probability ratio of sub-levels of N($^2$D). As discussed in
the earlier section, it is difficult to measure the [NI] ratio 
experimentally due to the N($^2$D) long lifetime  {($\sim$10
hrs)} while the collisional quenching in the terrestrial environment 
can alter the observed [NI] ratio significantly. Hence, the 
cometary spectroscopic observations are a new natural
way to explore atomic and molecular
properties that can not be determined in the laboratory.

\section{Summary and Conclusions}
\label{sec:sumary}
The recent observations of C/2016 R2  have shown that 
its coma has a peculiar composition compared to most comets
 observed so far. Several forbidden emission lines have been 
 detected in the optical, some for the first time, using 
 UVES mounted on the Very Large Telescope \citep{Opitom19}. 
 The observed atomic carbon,  oxygen, and nitrogen 
 forbidden emission transitions are 
 modelled under the frame-work of a coupled-chemistry 
 emission model, by incorporating the major production, 
 loss, and transport mechanisms of  { C($^1$D \& $^1$S)}, O($^1$D \& 
 $^1$S), and  N($^2$D) species. The observation of these 
 forbidden emissions  in a water depleted comet is a first 
 and unique case to study the photochemistry of the 
 metastable states and their emission processes in such a 
 cometary coma. The major conclusions drawn from both 
 observations and modelling  are as follows.

\begin{enumerate}
\item The major source of [OI], [CI], and 
[NI] emissions is found to be the photodissociation 
of CO$_2$, CO, and N$_2$, respectively.

\item The collisional quenching of all the metastable 
states, except for O($^1$S)  {and C($^1$S),} is 
significant for radial 
distances below 3 $\times$ 10$^3$ km. The collisional 
quenching of C($^1$D) and N($^2$D) occurs mainly by CO 
whereas it is by CO$_2$ for O($^1$D). 

\item In this water depleted coma, the observed 
[OI] G/R emission ratio is found to vary between 0.3 and 
0.18, which is smaller than the \cite{Festou81}   
value of 0.9 for a CO$_2$ coma by a factor of 3 or more. This 
lower value indicates that the observed [OI] G/R emission ratio 
 {is not a parameter} to identify the major parent 
source, neither
water nor CO$_2$. 

\item The observed  width of the green emission line is larger
 mainly because of the photodissociation of CO$_2$ that
 occurs by high energetic photons, which results in 
 O($^1$S) having large excess mean velocity. Since the 
photodissociation of CO$_2$ is the only source of the 
oxygen emission lines, the involvement of multiple species 
in producing the larger green line width can be discarded 
in this comet.

\item  By comparing the observed and modelled [CI]/[NI] 
emission ratios in this comet, we suggest that the 
photodissociation cross section of CO producing C($^1$D) 
should be reduced by a factor  {of 4.} 

\item Using a high molecular oxygen abundance (about 30\% 
with respect to CO), our modelled [OI] G/R ratio is 
consistent with the observations for radial distances below 10$^3$ 
km. Above this radial distance, the modelled values are 
higher by 40\% than the observations. This suggests that the
 photodissociation of
molecular oxygen  is a possible 
source of these emissions lines.

\item   Based on the observations 
and modelling of these emissions we suggest that in a 
water-depleted and CO and CO$_2$ rich cometary coma, which 
could be the case for comets observed at large 
($>$3 au) heliocentric distance as H$_2$O is not sublimating yet, 
 the observed [OI] emissions are mainly governed by CO$_2$.

\item  The modelling of the [OI] G/R ratio suggests 
that the ratio of dissociation frequencies of CO$_2$ 
producing O($^1$S) to O($^1$D) should be around 0.2. 
Similarly, the modelled [CI]/[NI] emission ratio suggests 
that mean branching fraction of CO for the production of 
C($^1$D)  should be smaller by a factor of 3 to 5 than the 
\cite{Huebner92} values. The observed emission ratios of 
metastable species  in the non-collisional region is a new 
way to constrain the  mean branching ratio of metastable 
species produced via photodissociative excitation. 

\item In spite of having a long radiative lifetime 
 {($\sim$10 hrs),} N($^2$D) is mainly controlled by transport rather 
than collisional quenching. Hence the observed [NI] ratio 
of 1.22, which is smaller than the terrestrial measurement by a
factor of 0.7, is mainly due to the intrinsic transition probability
ratio.

\item The simultaneous observation of both [CI] 8727 \AA\ 
and 9850 \AA\ emission lines allowed us to constrain the 
yield of CO photodissociation producing C($^1$S). It is 
about 1\% of the total absorption cross section, which 
was not previously determined in the laboratory.  This suggests that  
cometary spectroscopic observations serves as a natural 
laboratory to explore the atomic and molecular properties. 
\end{enumerate}

\begin{acknowledgements}
SR is supported by Department of Science and Technology 
(DST) with Innovation in
Science Pursuit for Inspired Research (INSPIRE) faculty 
award [Grant: DST/INSPIRE/04/2016/002687],
and he would like to thank Physical Research Laboratory for 
facilitating conducive research environment. DH and EJ are 
FNRS Senior Research Associates.  {The authors would like to
thank the 
anonymous reviewer for the valuable comments and suggestions
that  improved the manuscript.}
\end{acknowledgements}


\begin{thebibliography}{75}
	\expandafter\ifx\csname natexlab\endcsname\relax\def\natexlab#1{#1}\fi
	
	\bibitem[{{Altwegg} {et~al.}(2019){Altwegg}, {Balsiger}, \&
		{Fuselier}}]{Altwegg19}
	{Altwegg}, K., {Balsiger}, H., \& {Fuselier}, S.~A. 2019, \araa, 57, 113
	
	\bibitem[{Bhardwaj(1999)}]{Bhardwaj99a}
	Bhardwaj, A. 1999, J. Geophys. Res., 104, 1929
	
	\bibitem[{{Bhardwaj} \& {Haider}(2002)}]{Bhardwaj02}
	{Bhardwaj}, A. \& {Haider}, S.~A. 2002, Adv. Space Res., 29, 745
	
	\bibitem[{{Bhardwaj} \& {Raghuram}(2012)}]{Bhardwaj12}
	{Bhardwaj}, A. \& {Raghuram}, S. 2012, Astrophys. J., 748, 13
	
	\bibitem[{{Bieler} {et~al.}(2015){Bieler}, {Altwegg}, {Balsiger}, {Bar-Nun},
		{Berthelier}, {Bochsler}, {Briois}, {Calmonte}, {Combi}, {de Keyser}, 
		{van
			Dishoeck}, {Fiethe}, {Fuselier}, {Gasc}, {Gombosi}, {Hansen}, 
			{H{\"a}ssig},
		{J{\"a}ckel}, {Kopp}, {Korth}, {Le Roy}, {Mall}, {Maggiolo}, {Marty},
		{Mousis}, {Owen}, {R{\`e}me}, {Rubin}, {S{\'e}mon}, {Tzou}, {Waite}, 
		{Walsh},
		\& {Wurz}}]{Bieler15}
	{Bieler}, A., {Altwegg}, K., {Balsiger}, H., {et~al.} 2015, Nauture, 526, 
	678
	
	\bibitem[{{Biver} {et~al.}(2018){Biver}, {Bockel{\'e}e-Morvan}, {Paubert},
		{Moreno}, {Crovisier}, {Boissier}, {Bertrand}, {Boussier}, {Kugel}, 
		{McKay},
		{Dello Russo}, \& {DiSanti}}]{Biver18}
	{Biver}, N., {Bockel{\'e}e-Morvan}, D., {Paubert}, G., {et~al.} 2018, 
	Astron.
	\& Astrophys, 619, A127
	
	\bibitem[{{Cessateur} {et~al.}(2016{\natexlab{a}}){Cessateur}, {de Keyser},
		{Maggiolo}, {Gibbons}, {Gronoff}, {Gunell}, {Dhooghe}, {Loreau}, 
		{Vaeck}, \&
		{Altwegg}}]{Cessateur16a}
	{Cessateur}, G., {de Keyser}, J., {Maggiolo}, R., {et~al.} 
	2016{\natexlab{a}},
	Journal of Geophysical Research (Space Physics), 121, 804
	
	\bibitem[{{Cessateur} {et~al.}(2016{\natexlab{b}}){Cessateur}, {De Keyser},
		{Maggiolo}, {Rubin}, {Gronoff}, {Gibbons}, {Jehin}, {Dhooghe}, 
		{Gunell}, \&
		{Vaeck}}]{Cessateur16b}
	{Cessateur}, G., {De Keyser}, J., {Maggiolo}, R., {et~al.} 
	2016{\natexlab{b}},
	Mon. Not. R. Astron. Soc., 462, S116
	
	\bibitem[{{Cochran}(2008)}]{Cochran08}
	{Cochran}, A.~L. 2008, Icarus, 198, 181
	
	\bibitem[{{Cochran} \& {McKay}(2018)}]{Cochran18}
	{Cochran}, A.~L. \& {McKay}, A.~J. 2018, Astrophy. J. Lett., 854, L10
	
	\bibitem[{{Cochran} \& {Schleicher}(1993)}]{Cochran93}
	{Cochran}, A.~L. \& {Schleicher}, D.~G. 1993, \icarus, 105, 235
	
	\bibitem[{{de Le{\'o}n} {et~al.}(2019){de Le{\'o}n}, {Licandro},
		{Serra-Ricart}, {Cabrera-Lavers}, {Font Serra}, {Scarpa}, {de la Fuente
			Marcos}, \& {de la Fuente Marcos}}]{deleion19}
	{de Le{\'o}n}, J., {Licandro}, J., {Serra-Ricart}, M., {et~al.} 2019, 
	Research
	Notes of the American Astronomical Society, 3, 131
	
	\bibitem[{{de Val-Borro} {et~al.}(2018){de Val-Borro}, {Milam}, {Cordiner},
		{Charnley}, {Villanueva}, \& {Kuan}}]{Borro18}
	{de Val-Borro}, M., {Milam}, S.~N., {Cordiner}, M.~A., {et~al.} 2018, The
	Astronomer's Telegram, 11254
	
	\bibitem[{{Decock} {et~al.}(2013){Decock}, {Jehin}, {Hutsem{\'e}kers}, \&
		{Manfroid}}]{Decock13}
	{Decock}, A., {Jehin}, E., {Hutsem{\'e}kers}, D., \& {Manfroid}, J. 2013, 
	\aap,
	555, A34
	
	\bibitem[{{Decock} {et~al.}(2015){Decock}, {Jehin}, {Rousselot},
		{Hutsem{\'e}kers}, {Manfroid}, {Raghuram}, {Bhardwaj}, \&
		{Hubert}}]{Decock15}
	{Decock}, A., {Jehin}, E., {Rousselot}, P., {et~al.} 2015, Astron. \&
	Astrophys., 573, A1
	
	\bibitem[{{Delsemme} \& {Combi}(1976)}]{Delsemme76}
	{Delsemme}, A.~H. \& {Combi}, M.~R. 1976, Astrophys. J. Lett., 209, L149
	
	\bibitem[{Delsemme \& Combi(1979)}]{Delsemme79}
	Delsemme, A.~H. \& Combi, M.~R. 1979, Astrophys. J., 228, 330
	
	\bibitem[{{El-Qadi} \& {Stancil}(2013)}]{Qadi13}
	{El-Qadi}, W.~H. \& {Stancil}, P.~C. 2013, Astrophys. J., 779, 97
	
	\bibitem[{{Feldman} {et~al.}(2018){Feldman}, {A'Hearn}, Bertaux, Feaga, 
	Keeney,
		Knight, Noonan, Parker, Schindhelm, Steffl, Stern, Vervack, \&
		Weaver}]{Feldman18}
	{Feldman}, P.~D., {A'Hearn}, M.~F., Bertaux, J.-L., {et~al.} 2018, The
	Astronomical Journal, 155, 9
	
	\bibitem[{Feldman \& Brune(1976)}]{Feldman76}
	Feldman, P.~D. \& Brune, W.~H. 1976, Astrophys. J. Lett., 209, L45
	
	\bibitem[{Feldman {et~al.}(2004)Feldman, Cochran, \& Combi}]{Feldman04}
	Feldman, P.~D., Cochran, A.~L., \& Combi, M.~R. 2004, {Spectroscopic
		investigations of fragment species in the coma: Comets II} (M. C. 
		Festou, H.
	A. Weaver, \& H. U. Keller (Ed.)(Tucson: Univ. of Arizona)), 425--447
	
	\bibitem[{{Feldman} {et~al.}(1997){Feldman}, {Festou}, {Tozzi}, \&
		{Weaver}}]{Feldman97}
	{Feldman}, P.~D., {Festou}, M.~C., {Tozzi}, P., \& {Weaver}, H.~A. 1997,
	Astrophys. J., 475, 829
	
	\bibitem[{{Feldman} {et~al.}(1980){Feldman}, {Weaver}, {Festou}, {A'Hearn},
		{Jackson}, {Donn}, {Rahe}, {Smith}, \& {Benvenuti}}]{Feldman80}
	{Feldman}, P.~D., {Weaver}, H.~A., {Festou}, M., {et~al.} 1980, Nature, 286,
	132
	
	\bibitem[{Festou \& Feldman(1981)}]{Festou81}
	Festou, M.~C. \& Feldman, P.~D. 1981, Astron. Astrophys., 103, 154
	
	\bibitem[{Fink \& Johnson(1984)}]{Fink84}
	Fink, U. \& Johnson, J.~R. 1984, Astron. J., 89, 1565
	
	\bibitem[{{Fox}(1993)}]{Fox93}
	{Fox}, J.~L. 1993, J. Geophys. Res., 98, 3297
	
	\bibitem[{Furusho {et~al.}(2006)Furusho, Kawakitab, Fusec, \&
		Watanabe}]{Furusho06}
	Furusho, R., Kawakitab, H., Fusec, T., \& Watanabe, J. 2006, Adv. Space 
	Res.,
	9, 1983
	
	\bibitem[{Gao \& Ng(2019)}]{Gao19}
	Gao, H. \& Ng, C.-Y. 2019, Chinese Journal of Chemical Physics, 32, 23
	
	\bibitem[{Haser(1957)}]{Haser57}
	Haser, L. 1957, Bull. Acad. R Sci Liege, 43, 740
	
	\bibitem[{{Hernandez} \& {Turtle}(1969)}]{Hernandez69}
	{Hernandez}, G. \& {Turtle}, J.~P. 1969, Planetary and Space Science, 17, 
	675
	
	\bibitem[{Herron(1999)}]{Herron99}
	Herron, J.~T. 1999, Journal of Physical and Chemical Reference Data, 28, 
	1453
	
	\bibitem[{{H{\"o}lscher}(2015)}]{Alexander15}
	{H{\"o}lscher}, A. 2015, PhD thesis, Technische Universit\"{a}t Berlin,
	Fakult\"{a}t II - Mathematik und Naturwissenschaften
	
	\bibitem[{Huebner \& Carpenter(1979)}]{Huebner79}
	Huebner, W.~F. \& Carpenter, C.~W. 1979, Los Alamos Report, 8085
	
	\bibitem[{Huebner {et~al.}(1992)Huebner, Keady, \& Lyon}]{Huebner92}
	Huebner, W.~F., Keady, J.~J., \& Lyon, S.~P. 1992, Astrophys. Space Sci., 
	195,
	1
	
	\bibitem[{Huestis \& Slanger(2006)}]{Huestis06}
	Huestis, D.~L. \& Slanger, T.~G. 2006, American Astronomical Society, 38, 
	62.20
	
	\bibitem[{{Kanik} {et~al.}(2003){Kanik}, {Noren}, {Makarov}, {Vattipalle},
		{Ajello}, \& {Shemansky}}]{Kanik03}
	{Kanik}, I., {Noren}, C., {Makarov}, O.~P., {et~al.} 2003, J. Geophys. Res.,
	108, 5126
	
	\bibitem[{{Keeney} {et~al.}(2017){Keeney}, {Stern}, {A'Hearn}, {Bertaux},
		{Feaga}, {Feldman}, {Medina}, {Parker}, {Pineau}, {Schindhelm}, 
		{Steffl},
		{Versteeg}, \& {Weaver}}]{Keeney17}
	{Keeney}, B.~A., {Stern}, S.~A., {A'Hearn}, M.~F., {et~al.} 2017, Mon. Not. 
	R.
	Astron. Soc., 469, S158
	
	\bibitem[{Lawrence(1972)}]{Lawrence72a}
	Lawrence, G.~M. 1972, J. Chem. Phys., 57, 5616
	
	\bibitem[{{Lu} {et~al.}(2015){Lu}, {Chang}, {Benitez}, {Luo}, {Houria},
		{Ayari}, {Al Mogren}, {Hochlaf}, {Jackson}, \& {Ng}}]{Lu15}
	{Lu}, Z., {Chang}, Y.~C., {Benitez}, Y., {et~al.} 2015, Physical Chemistry
	Chemical Physics (Incorporating Faraday Transactions), 17, 11752
	
	\bibitem[{Magee-Sauer {et~al.}(1990)Magee-Sauer, Scherb, Roesler, \&
		Harlander}]{Magee90}
	Magee-Sauer, K., Scherb, F., Roesler, F.~L., \& Harlander, J. 1990, Icarus, 
	84,
	154
	
	\bibitem[{McElroy \& McConnell(1971)}]{Mcelroy71}
	McElroy, M.~B. \& McConnell, J.~C. 1971, J. Geophys. Res., 76, 6674
	
	\bibitem[{{McKay} {et~al.}(2019){McKay}, {DiSanti}, {Kelley}, {Knight},
		{Womack}, {Wierzchos}, {Harrington-Pinto}, {Bonev}, {Villanueva}, {Dello
			Russo}, {Cochran}, {Biver}, {Bauer}, {Vervack}, {Gibbs}, {Roth}, \&
		{Kawakita}}]{McKay19}
	{McKay}, A., {DiSanti}, M., {Kelley}, M., {et~al.} 2019, arXiv e-prints,
	arXiv:1907.07208
	
	\bibitem[{McKay {et~al.}(2012{\natexlab{a}})McKay, Chanover, Morgenthaler,
		Cochran, Harris, \& Russo}]{McKay12}
	McKay, A.~J., Chanover, N.~J., Morgenthaler, J.~P., {et~al.}
	2012{\natexlab{a}}, Icarus, 277
	
	\bibitem[{McKay {et~al.}(2012{\natexlab{b}})McKay, Chanover, Morgenthaler,
		Cochran, Harris, \& Russo}]{McKay12b}
	McKay, A.~J., Chanover, N.~J., Morgenthaler, J.~P., {et~al.}
	2012{\natexlab{b}}, Icarus
	
	\bibitem[{Morgenthaler {et~al.}(2001)Morgenthaler, Harris, Scherb, Anderson,
		Oliversen, Doane, Combi, Marconi, \& Smyth}]{Morgenthaler01}
	Morgenthaler, J.~P., Harris, W.~M., Scherb, F., {et~al.} 2001, Astrophys. 
	J,,
	563, 451
	
	\bibitem[{{Nussbaumer} \& {Rusca}(1979)}]{Nussbaumer79}
	{Nussbaumer}, H. \& {Rusca}, C. 1979, Astron. \& Astrophys., 72, 129
	
	\bibitem[{{Oliversen} {et~al.}(2002){Oliversen}, {Doane}, {Scherb}, 
	{Harris},
		\& {Morgenthaler}}]{Oliversen02}
	{Oliversen}, R.~J., {Doane}, N., {Scherb}, F., {Harris}, W.~M., \&
	{Morgenthaler}, J.~P. 2002, Astrophys. J., 581, 770
	
	\bibitem[{{Opitom} {et~al.}(2019){Opitom}, {Hutsem{\'e}kers}, {Jehin},
		{Rousselot}, {Pozuelos}, {Manfroid}, {Moulane}, {Gillon}, \&
		{Benkhaldoun}}]{Opitom19}
	{Opitom}, C., {Hutsem{\'e}kers}, D., {Jehin}, E., {et~al.} 2019, Astron. \&
	Astrophys., 624, A64
	
	\bibitem[{{Pan} {et~al.}(2011){Pan}, {Gao}, {Yang}, {Zhou}, {Ng}, \&
		{Jackson}}]{Pan11}
	{Pan}, Y., {Gao}, H., {Yang}, L., {et~al.} 2011, Journal of Chemical 
	Physics,
	135, 071101
	
	\bibitem[{{Raghuram} \& {Bhardwaj}(2013)}]{Raghuram13}
	{Raghuram}, S. \& {Bhardwaj}, A. 2013, Icarus, 223, 91
	
	\bibitem[{{Raghuram} \& {Bhardwaj}(2014)}]{Raghuram14}
	{Raghuram}, S. \& {Bhardwaj}, A. 2014, Astron. Astrophys., 566, A134
	
	\bibitem[{{Raghuram} {et~al.}(2016){Raghuram}, {Bhardwaj}, \&
		{Galand}}]{Raghuram16}
	{Raghuram}, S., {Bhardwaj}, A., \& {Galand}, M. 2016, Astrophys. J., 818, 
	102
	
	\bibitem[{{Rodgers} {et~al.}(2004){Rodgers}, {Charnley}, {Huebner}, \&
		{Boice}}]{Rodgers04}
	{Rodgers}, S.~D., {Charnley}, S.~B., {Huebner}, W.~F., \& {Boice}, D.~C. 
	2004,
	{Physical Processes and Chemical Reactions in Cometary Comae: Comets II}, 
	ed.
	{Festou, M.~C., Keller, H.~U., \& Weaver, H.~A.}, 505--522
	
	\bibitem[{{Rubin} {et~al.}(2015){Rubin}, {Altwegg}, {van Dishoeck}, \&
		{Schwehm}}]{Rubin15}
	{Rubin}, M., {Altwegg}, K., {van Dishoeck}, E.~F., \& {Schwehm}, G. 2015,
	Astrophys. J., 815, L11
	
	\bibitem[{Saxena {et~al.}(2002)Saxena, Bhatnagar, \& Singh}]{Saxena02}
	Saxena, P.~P., Bhatnagar, S., \& Singh, M. 2002, Mon. Not. R. Astron. Soc.,
	334, 563
	
	\bibitem[{{Schmidt} {et~al.}(2013){Schmidt}, {Johnson}, \&
		{Schinke}}]{Schmidt13}
	{Schmidt}, J.~A., {Johnson}, M.~S., \& {Schinke}, R. 2013, Proceedings of 
	the
	National Academy of Science, 110, 17691
	
	\bibitem[{{Schofield}(1979)}]{Schofield79}
	{Schofield}, K. 1979, Journal of Physical and Chemical Reference Data, 8, 
	723
	
	\bibitem[{Schultz {et~al.}(1992)Schultz, Li, Scherb, \& Roesler}]{Schultz92}
	Schultz, D., Li, G. S.~H., Scherb, F., \& Roesler, F.~L. 1992, Icarus, 96, 
	190
	
	\bibitem[{{Sharpee} {et~al.}(2005){Sharpee}, {Slanger}, {Cosby}, \&
		{Huestis}}]{Sharpee05}
	{Sharpee}, B.~D., {Slanger}, T.~G., {Cosby}, P.~C., \& {Huestis}, D.~L. 
	2005,
	Geophys. Res. Lett., 32, L12106
	
	\bibitem[{{Shi} {et~al.}(2018){Shi}, {Gao}, {Yin}, {Chang}, {Wiens}, 
	{Jackson},
		\& {Ng}}]{Shi18}
	{Shi}, X., {Gao}, H., {Yin}, Q.-Z., {et~al.} 2018, Journal of Physical
	Chemistry A, 122, 8136
	
	\bibitem[{{Singh} {et~al.}(1991){Singh}, {D'Ealmeida}, \& 
	{Huebner}}]{Singh91}
	{Singh}, P.~D., {D'Ealmeida}, A.~A., \& {Huebner}, W.~F. 1991, Icarus, 90, 
	74
	
	\bibitem[{{Sivjee} {et~al.}(1981){Sivjee}, {Deehr}, \& 
	{Henriksen}}]{Sivjee81}
	{Sivjee}, G.~G., {Deehr}, C.~S., \& {Henriksen}, K. 1981, J. Geophys. Res., 
	86,
	1581
	
	\bibitem[{{Slanger} {et~al.}(1977){Slanger}, {Sharpless}, \&
		{Black}}]{Slanger77}
	{Slanger}, T.~G., {Sharpless}, R.~L., \& {Black}, G. 1977, Journal of 
	Chemical
	Physics, 67, 5317
	
	\bibitem[{Smith {et~al.}(1980)Smith, Stecher, \& Casswell}]{Smith80}
	Smith, A.~M., Stecher, T.~P., \& Casswell, L. 1980, Astrophys. J., 242, 402
	
	\bibitem[{{Song} {et~al.}(2014){Song}, {Gao}, {Chang}, {Lu}, {Ng}, \&
		{Jackson}}]{Song14}
	{Song}, Y., {Gao}, H., {Chang}, Y.~C., {et~al.} 2014, Physical Chemistry
	Chemical Physics (Incorporating Faraday Transactions), 16, 563
	
	\bibitem[{{Song} {et~al.}(2016){Song}, {Gao}, {Chung Chang}, 
	{Hammout{\'e}ne},
		{Ndome}, {Hochlaf}, {Jackson}, \& {Ng}}]{Song16}
	{Song}, Y., {Gao}, H., {Chung Chang}, Y., {et~al.} 2016, Astrophys. J., 
	819, 23
	
	\bibitem[{Sutradhar {et~al.}(2017)Sutradhar, Samanta, Samanta, \&
		Reisler}]{Sutradhar17}
	Sutradhar, S., Samanta, B.~R., Samanta, A.~K., \& Reisler, H. 2017, The 
	Journal
	of Chemical Physics, 147, 013916
	
	\bibitem[{{Tabata} {et~al.}(2006){Tabata}, {Shirai}, M., \& 
	{Kubo}}]{Tabata06}
	{Tabata}, T., {Shirai}, T., M., S., \& {Kubo}, H. 2006, Atomic Data and 
	Nuclear
	Data Tables, 92, 375
	
	\bibitem[{{Tachiev} \& {Froese Fischer}(2002)}]{Tachiev02}
	{Tachiev}, G.~I. \& {Froese Fischer}, C. 2002, \aap, 385, 716
	
	\bibitem[{Tozzi {et~al.}(1998)Tozzi, Feldman, \& Festou}]{Tozzi98}
	Tozzi, G.~P., Feldman, P.~D., \& Festou, M.~C. 1998, Astron. Astrophys., 
	330,
	753
	
	\bibitem[{{Wierzchos} \& {Womack}(2018)}]{Wierzchos18}
	{Wierzchos}, K. \& {Womack}, M. 2018, Astron. J., 156, 34
	
	\bibitem[{{Wiese} \& {Fuhr}(2007)}]{Wiese07}
	{Wiese}, W.~L. \& {Fuhr}, J.~R. 2007, Journal of Physical and Chemical
	Reference Data, 36, 1287
	
	\bibitem[{Wiese \& Fuhr(2009)}]{Wiese09}
	Wiese, W.~L. \& Fuhr, J.~R. 2009, Journal of Physical and Chemical Reference
	Data, 38, 565
	
	\bibitem[{Wiese {et~al.}(1996)Wiese, Fuhr, \& Deters}]{Wiese96}
	Wiese, W.~L., Fuhr, J.~R., \& Deters, T.~M. 1996, {Atomic transition
		probabilities of carbon, nitrogen, and oxygen: A critical data 
		compilation}
	(Am. Chem. Soc., Washington, D. C.)
	
	\bibitem[{{Wu} \& {Chen}(1993)}]{Wu93}
	{Wu}, C.~Y.~R. \& {Chen}, F.~Z. 1993, J. Geophys. Res, 98, 7415
	
\end{thebibliography}
\end{document}